%% file: main.tex
\documentclass[journal]{./lib/IEEEtran}
\usepackage{cite,threeparttable,stfloats,amssymb,times,placeins,amsmath,multirow,booktabs,array,bm}
\usepackage[T1]{fontenc} 
\usepackage[caption=false,font=footnotesize,labelfont=rm,textfont=rm]{subfig}
\usepackage{xcolor, soul} 
\usepackage{colortbl}	
\sethlcolor{lime} 

\usepackage{ifpdf}
\ifpdf
\usepackage[pdftex]{graphicx}      
\DeclareGraphicsExtensions{.pdf,.png,.jpg,.mps}
\usepackage[pdftex,                
bookmarks=true,
bookmarksopen=true,%
bookmarksnumbered=false,
breaklinks=true,
pdfborder={0 0 0},					
colorlinks,
linkcolor=blue,
citecolor=blue,
anchorcolor=blue,
]{hyperref}
\else
\usepackage{graphicx}              
\DeclareGraphicsExtensions{.eps}
\usepackage[ps2pdf,                
bookmarks=true,
bookmarksnumbered=true,
breaklinks=true,
pdfborder={0 0 10}]{hyperref}
\fi

\graphicspath{{figures/}}

\hypersetup{%
pdfstartview={FitH},   
}%

\newcommand{\revision}{\textcolor{.}}

\newcommand{\sectionname}{Section~}
\renewcommand{\figurename}{Fig.~}
\renewcommand{\tablename}{Table~}

\begin{document}

\title{A Timing-Robust Frequency-Interleaved ADC with Embedded Analog Fourier Transform}

\author{Xingchen~Chao,
        Yunqiang~Xu,
        Qiang~Yu,
        Christian~Adam,
        and~Qiang~Li,
        
\thanks{Xingchen Chao, Yunqiang Xu and Qiang Yu are with Institute of Integrated Circuits and Systems, University of Electronic Science and Technology of China (UESTC), Chengdu 610054, China.

Christian Adam is with Hamburg University of Technology (TUHH), Hamburg 21073, Germany.

Qiang Li was with University of Electronic Science and Technology of China. He is currently with Hamburg University of Technology (TUHH), Hamburg 21073, Germany (qiang.li@tuhh.de).}}


\maketitle

\begin{abstract}
Timing errors represent a fundamental challenge in high-speed analog-to-digital converters (ADCs).
This paper presents a frequency-interleaved architecture with inherent robustness to timing errors.
The frequency interleaving is achieved by an embedded analog Fourier transform (AFT), allowing accurate channel spacing. 
The charge-based passive AFT is able to share the capacitor array with the capacitive DAC of the subsequent data conversion. 
A circuit implementation for the complex-valued irrational coefficients in the AFT matrix is proposed, which eliminates the device-ratio-induced systematic errors. 
Measured at 10b 13GS/s, the calibration-free ADC achieves 45.8-dB SNDR and 56.1-dB SFDR at Nyquist.
The ADC consumes 26.4mA from a single 1V supply, resulting in a Walden FoM of 12.7 fJ/conv.-step.


\end{abstract}


\IEEEpeerreviewmaketitle

\input{sections/1_introduction}
\input{sections/2_principle_and_error_analysis}

\input{sections/3_implementation_of_the_irrational_twiddle_factor}

\input{sections/4_circuit_implementation}
\input{sections/5_measurement_results}

\input{sections/6_conclusion}


\bibliographystyle{IEEEtran}
\bibliography{./lib/Reference}
\vspace{0cm}
\begin{IEEEbiography}[{\includegraphics[width=1in,height=1.25in,clip,keepaspectratio]{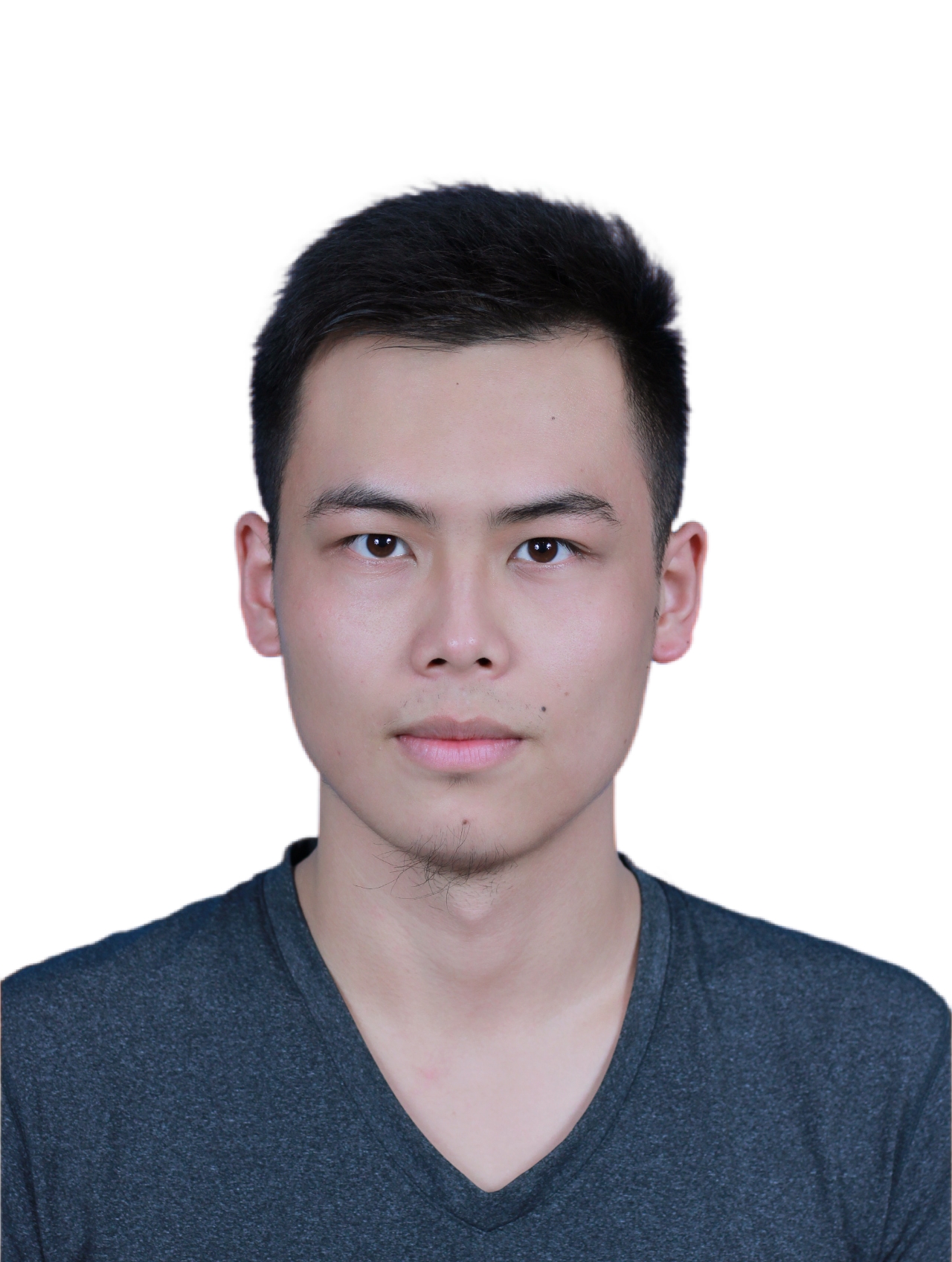}}]{Xingchen Chao}
	received the B.S. degree in Microelectronics Science and Engineering from the University of Electronic Science and Technology of China, Chengdu, China in 2021, where he is currently pursuing the Ph.D degree in the Institute of Integrated Circuits and Systems. 
	
	His current research interests include high-speed analog-to-digital converters and receiver architectures for high-speed wireline and wireless communication systems.
\end{IEEEbiography}

\vspace{0cm}
\begin{IEEEbiography}[{\includegraphics[width=1in,height=1.25in,clip,keepaspectratio]{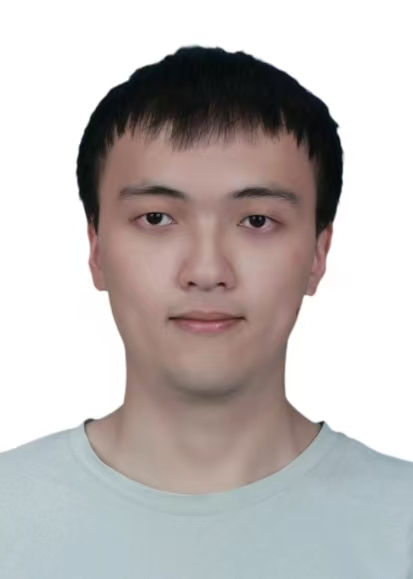}}]{Yunqiang Xu}
	received the B.S. degree in Microelectronics Science and Engineering from the University of Electronic Science and Technology of China, Chengdu, China in 2022, where he is currently pursuing the M.S. degree in the Institute of Integrated Circuits and Systems. 
	
	His research interests include high-speed ADC and communication systems.
\end{IEEEbiography}

\vspace{0cm}
\begin{IEEEbiography}[{\includegraphics[width=1in,height=1.25in,clip,keepaspectratio]{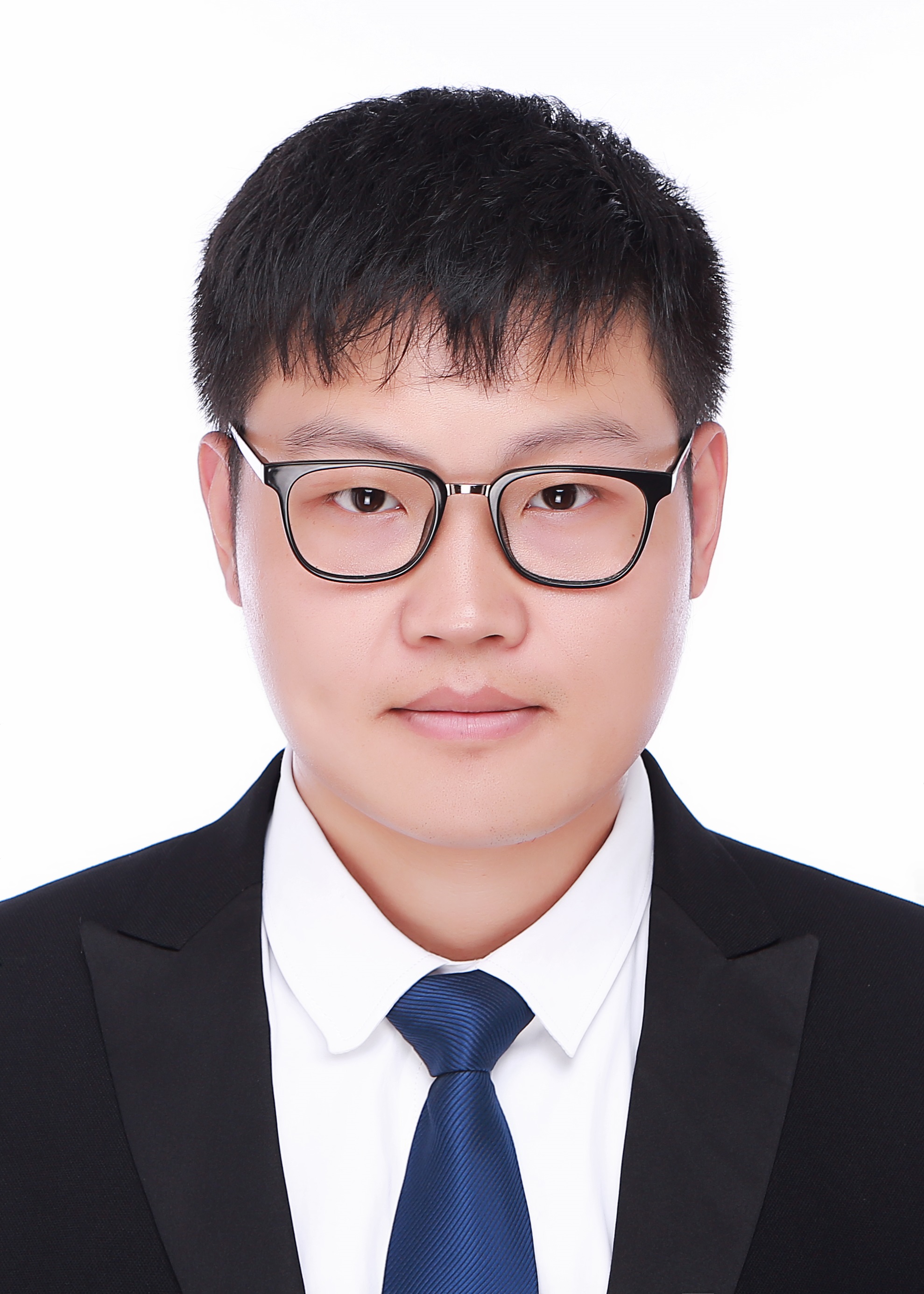}}]{Qiang Yu}
	received the B.S. and M.S. degrees in Microelectronics from the University of Electronic Science and Technology of China, Chengdu, China, in 2014 and 2019, respectively.
	He is currently pursuing the Ph.D. degree with the Institute of Integrated Circuits and Systems, University of Electronic Science and Technology, Chengdu, China. 
	
	His research interests include high-speed analog-to-digital converters and digitally assisted-analog design.
\end{IEEEbiography}

\vspace{0cm}
\begin{IEEEbiography}[{\includegraphics[width=1in,height=1.25in,clip,keepaspectratio]{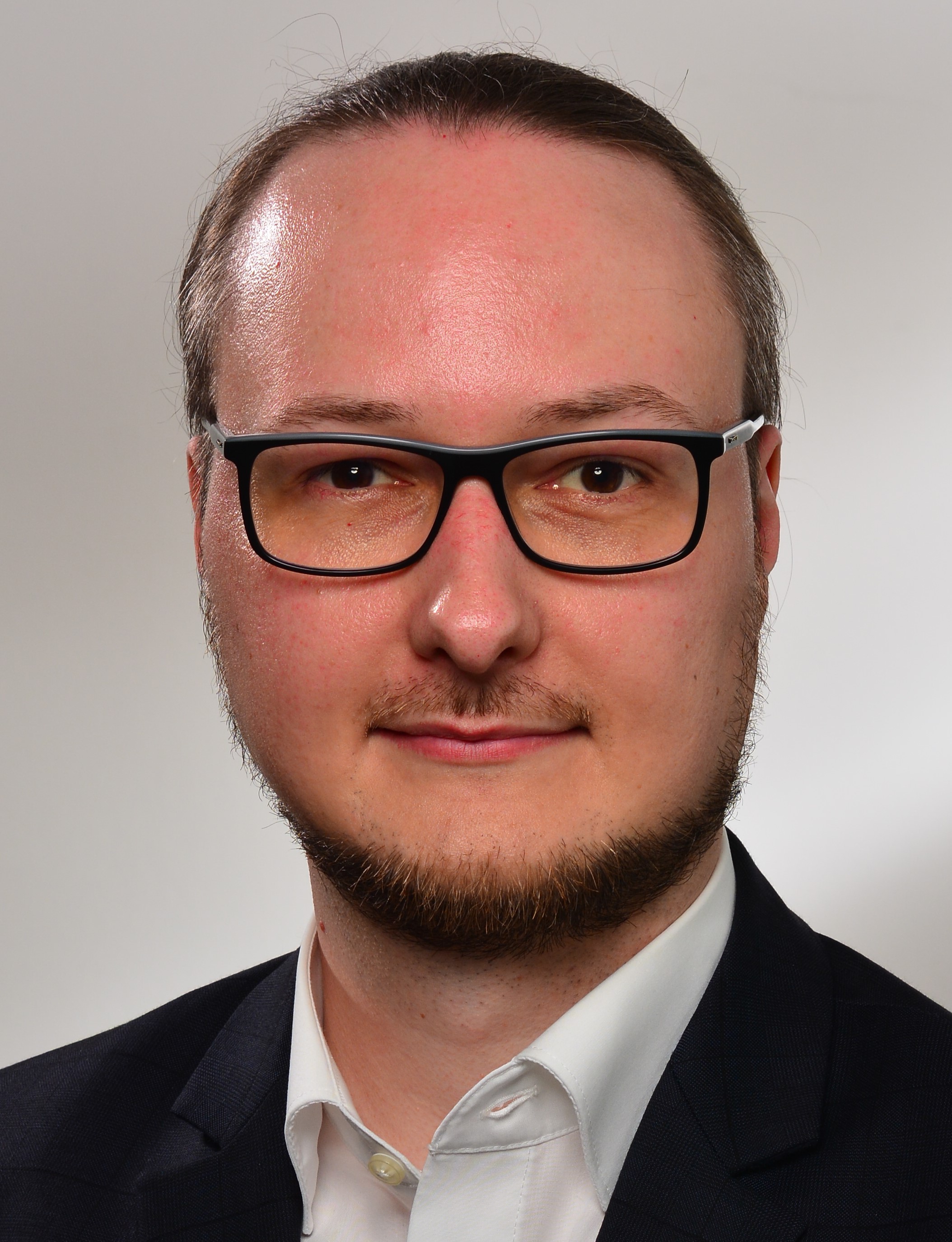}}]{Christian Adam}
	received the B.Sc. and M.Sc. degrees in electrical engineering from the Hamburg University of Technology (TUHH), Hamburg, Germany, in 2014 and 2017, where he also defended his Ph.D. in electrical engineering in 2025. 
	
	From 2017 to 2025 he was working on the development of medical devices for fracture treatment as an R\&D engineer with Innovations Medical and other companies. From 2019 to 2025 he also was a research associate at the Institute of Integrated Circuits and Systems at TUHH, where he is a senior engineer since 2025. His research interests include sensor interfaces and wireless communication for biomedical systems, and mixed-signal integrated circuits for signal processing.
\end{IEEEbiography}

\vspace{0cm}
\begin{IEEEbiography}[{\includegraphics[width=1in,height=1.25in,clip,keepaspectratio]{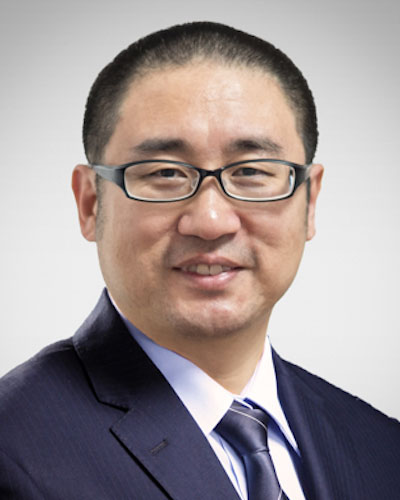}}]{Qiang Li}
	received the B.S. degree in Electrical Engineering from the Huazhong University of Science and Technology (HUST), Wuhan, China and the Ph.D. in Electronic Engineering from the Nanyang Technological University (NTU), Singapore, in 2001 and 2007, respectively. 
	
	He has been an engineer in Singapore before joining the University of Electronic Science and Technology of China (UESTC), Chengdu, in 2009, where he was a full Professor and Founding Head of the UESTC Institute of Integrated Circuits and Systems. Since 2025, he has been with the Hamburg University of Technology (TUHH), Germany, where he is a Professor and Head of the Institute of Integrated Circuits and Systems. His research interests include low-voltage and low-power analog/RF circuits, data converters, mixed-signal circuits for biomedical and sensor interfaces, and analog computing.
	
	Prof. Li was a Distinguished Lecturer of the IEEE Solid-State Circuits Society (SSCS) and on the Board of Governors (BoG) of IEEE Circuits and Systems Society (CASS). He served/serves as a member of Technical Program Committees for ISSCC, CICC, ASSCC, ESSCIRC, and as the TPC Chair of APCCAS’2018. He was the Founding Chair of ICAC Workshop since 2019, the premier forum for IC advances in China.
\end{IEEEbiography}

\end{document}

%% file: sections/1_introduction.tex
\section{Introduction}\label{sec:inctroduction}

\IEEEPARstart{T}{he} evolution of data rates in wireless and wireline systems has been continuously driving the development of high-speed analog-to-digital converters (ADCs).  
The trend toward multi-band operation and agile carrier aggregation in emerging mm-wave and sub-THz standards imposes unprecedented performance demands on wideband ADCs operating at tens of GS/s \cite{RFIC2022-Dascurcu,JSSC2019-Pang,RFIC2024-Hamani}.
Modern DSP-based SerDes for wireline and optical links deliver multi-hundred-Gbps throughput, driving the need for ultra-high-speed ADCs operating at tens to hundreds of GS/s\cite{ISSCC2024-Wang,ISSCC2024-Li,ISSCC2025-Chen}.

\begin{figure} [!t]
	\centering 
	{
		\includegraphics[width = 0.9\linewidth]{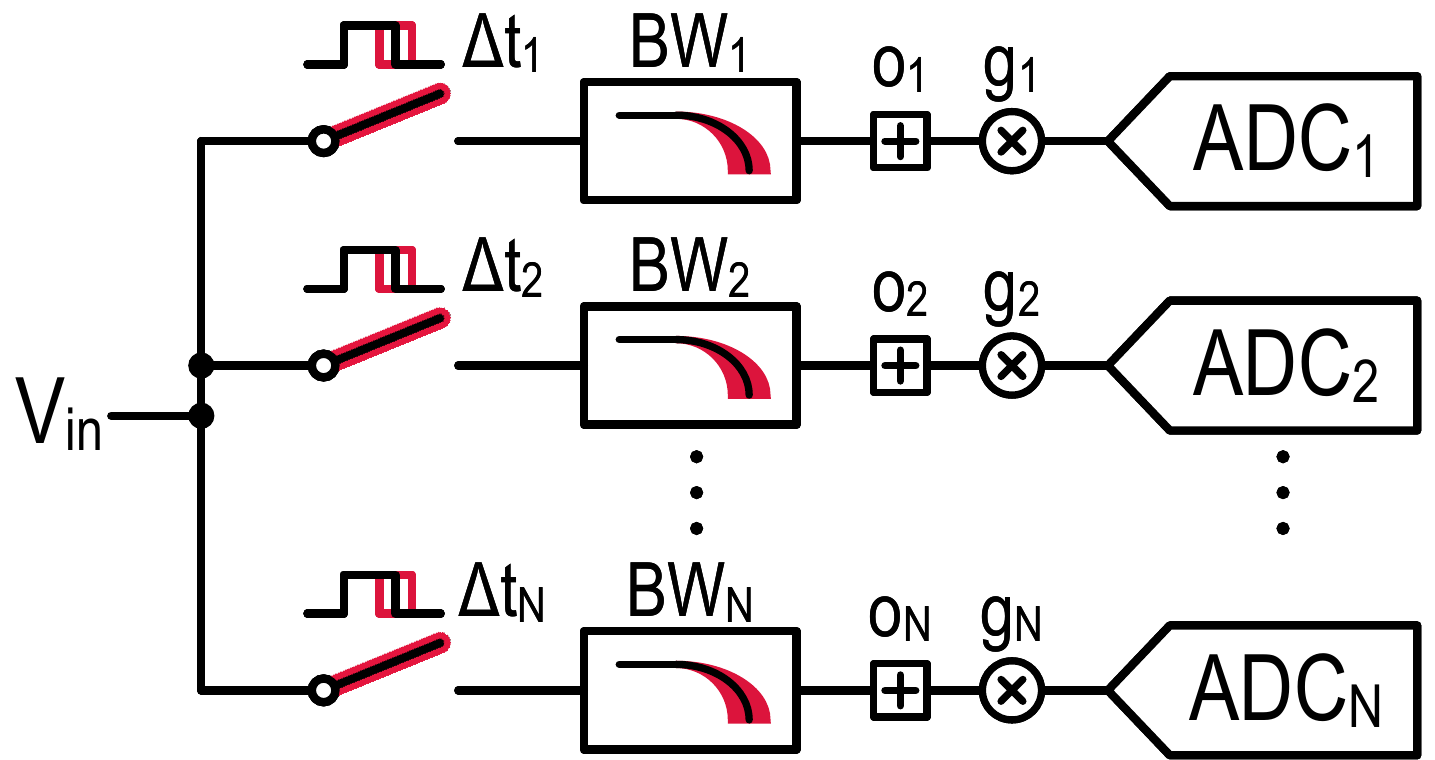}
	} 
		\caption{Time-interleaved ADC architecture with channel mismatches.} 
	\label{fig-1.1} 
\end{figure}

Time interleaving (TI) has become the de facto architecture for state-of-the-art high-speed ADCs.
As illustrated in \figurename\ref{fig-1.1}, a TI ADC consists of multiple parallel sub-ADCs, each sampling the input signal at an interleaved time instant, and the quantized outputs are recombined in the digital domain.
A fundamental challenge in TI ADC architectures lies in the inter-channel mismatches. 
Static mismatches, i.e., offset and gain mismatches, are relatively straightforward to mitigate as they remain independent of the input frequency \cite{IEEETrans2001-Kurosawa,JSSC2002-Jamal}.
In contrast, timing errors, including clock skew and tracking bandwidth mismatch, are more challenging to correct because of their frequency-dependent nature, and becoming increasingly severe at higher sampling rates and larger interleaving factors \cite{IEEETrans2001-Kurosawa,JSSC2013-Razavi}.


Calibration techniques for timing errors have been studied extensively in recent years. 
The errors due to timing skew can be estimated in the digital domain through three main approaches: employing additional reference channels \cite{JSSC2013-Stepanovic,JSSC2020-Guo}, extracting signal statistical properties such as autocorrelation and autocovariance \cite{JSSC2002-Jamal,JSSC2013-Razavi,JSSC2025-Gu}, or injecting global or local test signals \cite{ISSCC2023-Wang,ISSCC2024-Cao,ISSCC2024-Tao}. 
With the estimated skew, the correction can be implemented by tuning the delay elements in the analog domain \cite{JSSC2013-Razavi,JSSC2025-Gu,ISSCC2023-Wang,ISSCC2024-Cao,ISSCC2024-Tao,JSSC2018-Kull} or through adaptive filtering in the digital domain \cite{JSSC2002-Jamal,JSSC2013-Stepanovic,JSSC2020-Guo}. 
Calibration of tracking bandwidth mismatch has been discussed to a rather limited extent.
Reported approaches include the tuning of sampling capacitors \cite{VLSI2018-Kull} and the digital equalization of channel frequency responses \cite{TCAS1-Tsai}.
Mitigating timing errors in a TI system incurs considerable design overhead and hardware complexity.
Furthermore, the convergence and accuracy of the estimation algorithms are highly contingent upon the statistical properties of the input signal, often requiring wide-sense stationarity (WSS) and sufficient frequency excitation.

\begin{figure} [!t]	
	\centering
	{ 
		\includegraphics[width = 0.9\linewidth]{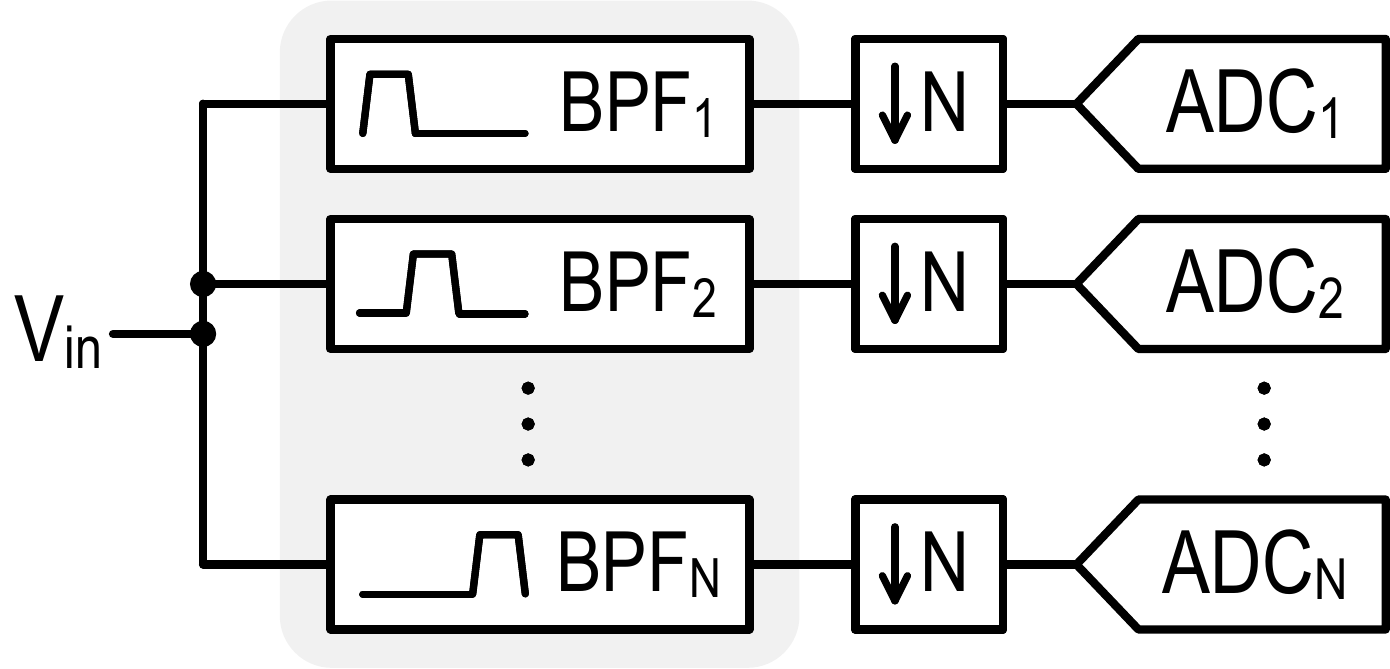} 
	}
	\caption{Frequency interleaving with hybrid filter banks (HFB).} 
	\label{fig-1.2} 
\end{figure}

Theoretical analysis demonstrates that frequency interleaving (FI) offers inherent robustness against analog mismatches, including timing errors \cite{IEEE1998}. 
\figurename\ref{fig-1.2} illustrates the FI architecture based on hybrid filter banks (HFB), which  decomposes the wideband input signal into multiple narrowband channels.
These channelized signals are subsequently downconverted to baseband and digitized by the sub-ADCs.
Compared with TI architectures, each sub-ADC processes only a fraction of the total bandwidth, thereby drastically reducing the sensitivity to frequency-dependent timing errors.
However, the implementation of HFB is impractical since each sub-channel works at a unique center frequency.
All analog filters and mixers must be individually customized, incurring excessive design complexity as well as hardware overhead. 
More importantly, the matching of frequency responses across these non-uniform components is intractable. 

Mathematically, the Fourier transform (FT) provides an ideal spectral decomposition, which has been explored for channelization in both receiver front-ends \cite{JSSC2013-Sadhu} and ADCs \cite{CICC2015-Shin}.
Previous research has theoretically demonstrated that a frequency-interleaved sampling system can be achieved via the analog implementation of the Fourier transform (AFT), which exhibits  substantial robustness to timing errors\cite{ISCAS2023-Chao}.
However, a frequency-interleaved ADC with compelling performance and timing robustness has not yet been demonstrated.


This work presents a 10-bit 13-GS/s frequency-interleaved ADC that eliminates the need for timing-error calibration, validating its inherent timing robustness.
The FI is achieved via an 8-point charge-based AFT which is embedded within the capacitive DACs (CDAC) of the subsequent sub-ADCs. 
The implementation accuracy is fundamentally limited by the complex-valued irrational coefficients in the AFT matrix, where device-ratio-based implementations lead to inevitable systematic errors.
In this work, a lossless implementation of the irrational coefficients is proposed. 
Fabricated in a 28-nm CMOS, the prototype ADC achieves 45.8-dB SNDR and 56.1-dB SFDR at a Nyquist input.
Under a single 1-V supply, the ADC consumes 26.4 mW, yielding a Walden figure-of-merit (FoM) of 12.7 fJ/conv.-step.

This work was first introduced in \cite{CICC2025-Chao}.
The complete architectural study with detailed analysis and design considerations is presented here.
\sectionname \ref{sec:principle_and_error_analysis} studies the AFT-based FI architecture and its inherent timing robustness.   
\sectionname \ref{sec:implementation_of_the_irrational_twiddle_factor} introduces the proposed circuit implementation of the complex-valued irrational coefficients. 
\sectionname \ref{sec:circuit_implementation} discusses the design considerations and implementation details.
\sectionname \ref{sec:measurement_results} shows the measurement results. 
This paper is concluded in \sectionname \ref{sec:conclusion}.

%% file: sections/2_principle_and_error_analysis.tex
\section{AFT-based Frequency Interleaving: Architecture and Timing Robustness}\label{sec:principle_and_error_analysis}

\begin{figure}[t]
	\centering 
	\subfloat[]	
	{
		\label{fig-2.1a} 
		\includegraphics[width = 0.7\linewidth]{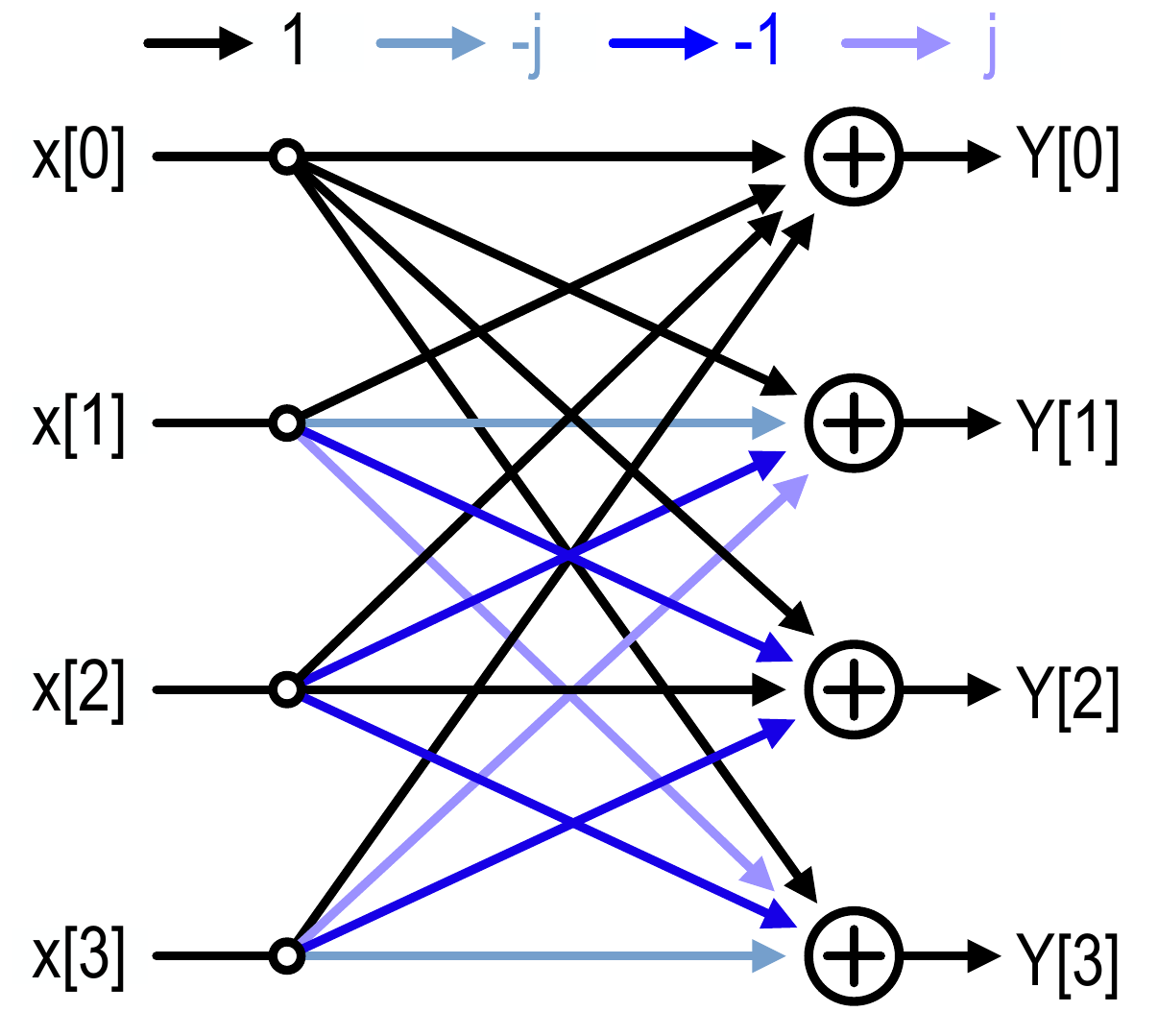}
	} 
	\\
	\subfloat[] 
	{ 
		\label{fig-2.1b} 
		\includegraphics[width = 0.85\linewidth]{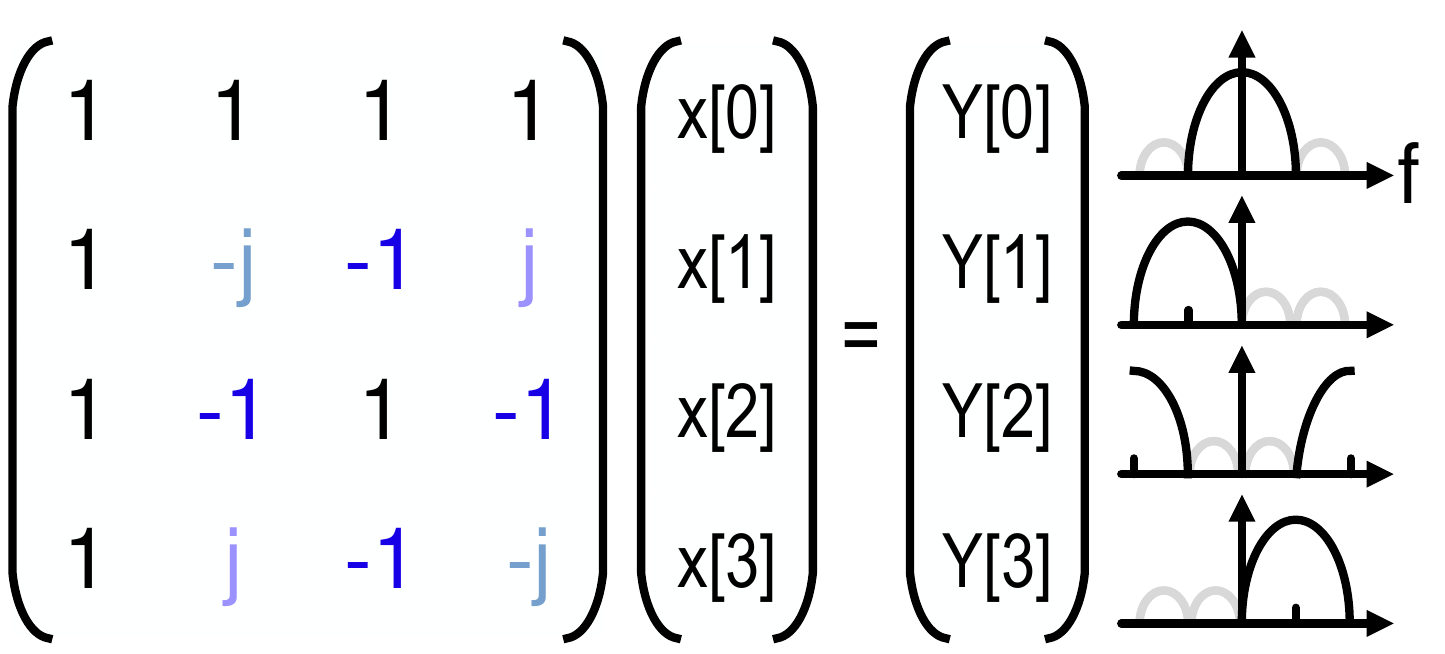} 
	}
	\caption{Conceptual illustration of (a) a 4-point AFT algorithm with (b) its matrix operation and spectral channelization.} 
	\label{fig-2.1} 
\end{figure}


\figurename\ref{fig-2.1}\subref{fig-2.1a} gives a conceptual illustration of a 4-point AFT-based FI system.
The mathematical foundation of AFT is the discrete Fourier transform (DFT), which can be written as
\begin{equation}\label{eq2.1}
	Y\left [ k \right ]= \sum_{n=0}^{N-1}x[n]e^{-j\frac{2\pi}{N}kn} , \quad k=0,1,...,N-1
\end{equation}
where $N$ denotes the number of DFT points, $x\left [ n \right ]$ is the $n$-th discrete-time input sample, and the exponential term represents the twiddle factor. By exploiting the orthogonality of the discrete Fourier basis, the frequency-domain output $Y\left [ k \right ]$ is derived from $x\left [ n \right ]$.
In \figurename\ref{fig-2.1}, the arrows represent the twiddle factors, with each color corresponding to a specific coefficient value.

\figurename\ref{fig-2.1}\subref{fig-2.1b} shows the matrix operation of the 4-point AFT.
Taking $Y\left [ 1 \right ]$ as an example, which can be expressed as
\begin{equation}\label{eq2.2}
         Y\left [ 1 \right ]= 1\times x\left [ 0 \right ]- j\times x\left [ 1 \right ]- x\left [ 2 \right ]+  j\times x\left [ 3 \right ].
\end{equation}
This equation can be interpreted as a weighted sum of the input samples $x[0]$, $x[1]$, $x[2]$, and $x[3]$, with the respective weights $1$, $-j$, $-1$, and $j$. 
This operation is equivalent to a 4-tap finite impulse response (FIR) filter, whose frequency response is illustrated on the right side of \figurename\ref{fig-2.1}\subref{fig-2.1b}.
From this perspective, each channel functions as a complex FIR filter with a distinct center frequency, and its tap coefficients correspond to the twiddle factors of the respective channel.
As a result, the AFT system enables broadband spectral channelization over the frequency range from $-f_{\mathrm{s}}/2$ to $f_{\mathrm{s}}/2$, where $f_{\mathrm{s}}$ denotes the sampling frequency of $x\left [ n \right ]$.

\begin{figure}[!t]
	\centering
	\includegraphics[width = 0.99\linewidth]{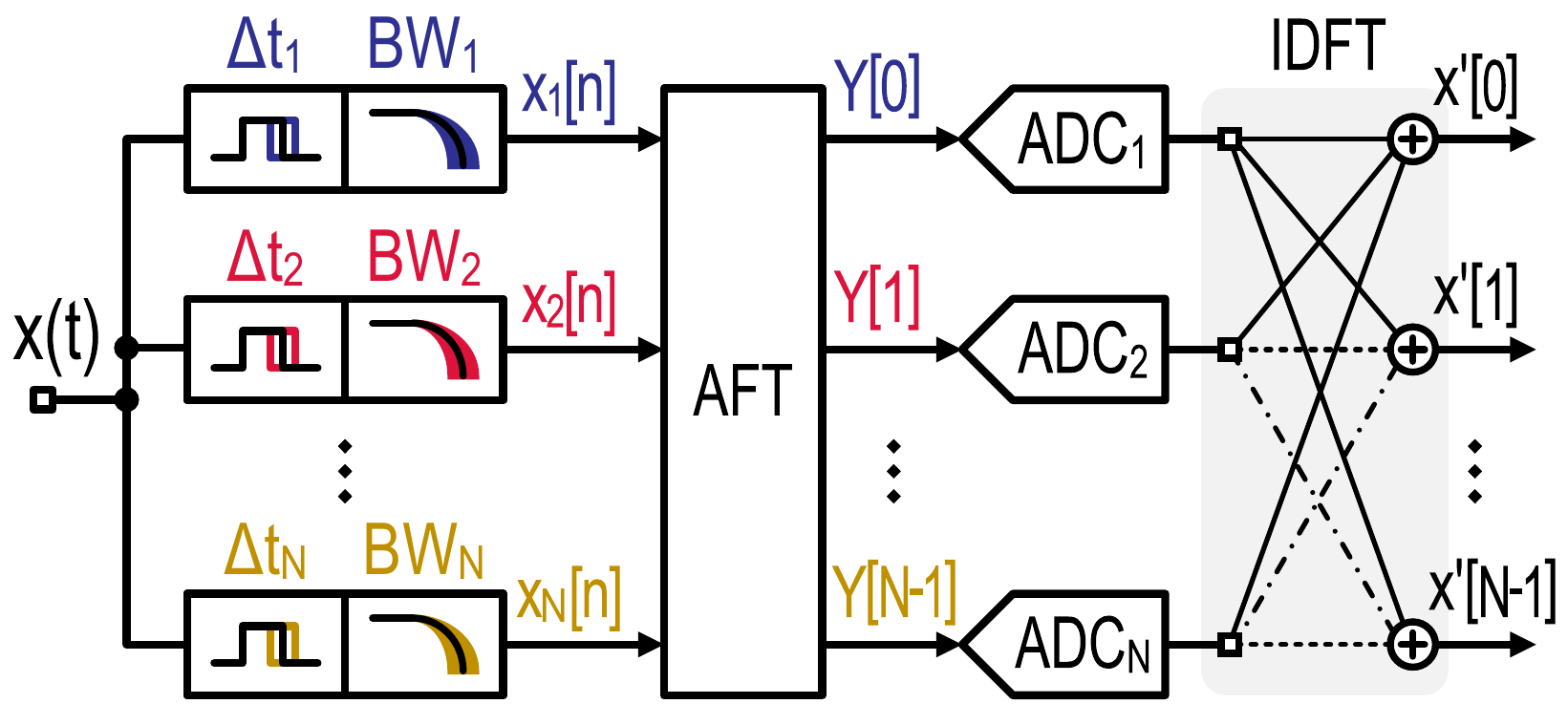}
	\caption{Architecture of the proposed AFT-based FI system with timing errors.}
	\label{fig-2.2}
\end{figure}

\figurename\ref{fig-2.2} shows the block diagram of the proposed AFT-based FI system with timing errors, where $\Delta t_{i}$ and $BW_{i}$ denote the timing skew and the tracking bandwidth of the $i$-th ($i=1, 2, \cdots, N$) channel.
$Y\left [ k \right ]$ ($k=0, 1, \cdots, N-1$) represents the frequency-domain components generated by the AFT, and $\mathsf{x'}[m]$ ($m=0, 1, \cdots, N-1$) denotes the time-domain signal reconstructed via an inverse discrete Fourier transform (IDFT) in the digital backend.
Analogous to the DFT, the IDFT reconstructs the time-domain signal by performing a weighted summation of the channelized spectral components via the IDFT coefficient matrix.

\begin{figure*}[!hb]
	\centering 
	\subfloat[]	
	{
		\label{fig-2.3a} 
		\includegraphics[width = 0.48\linewidth]{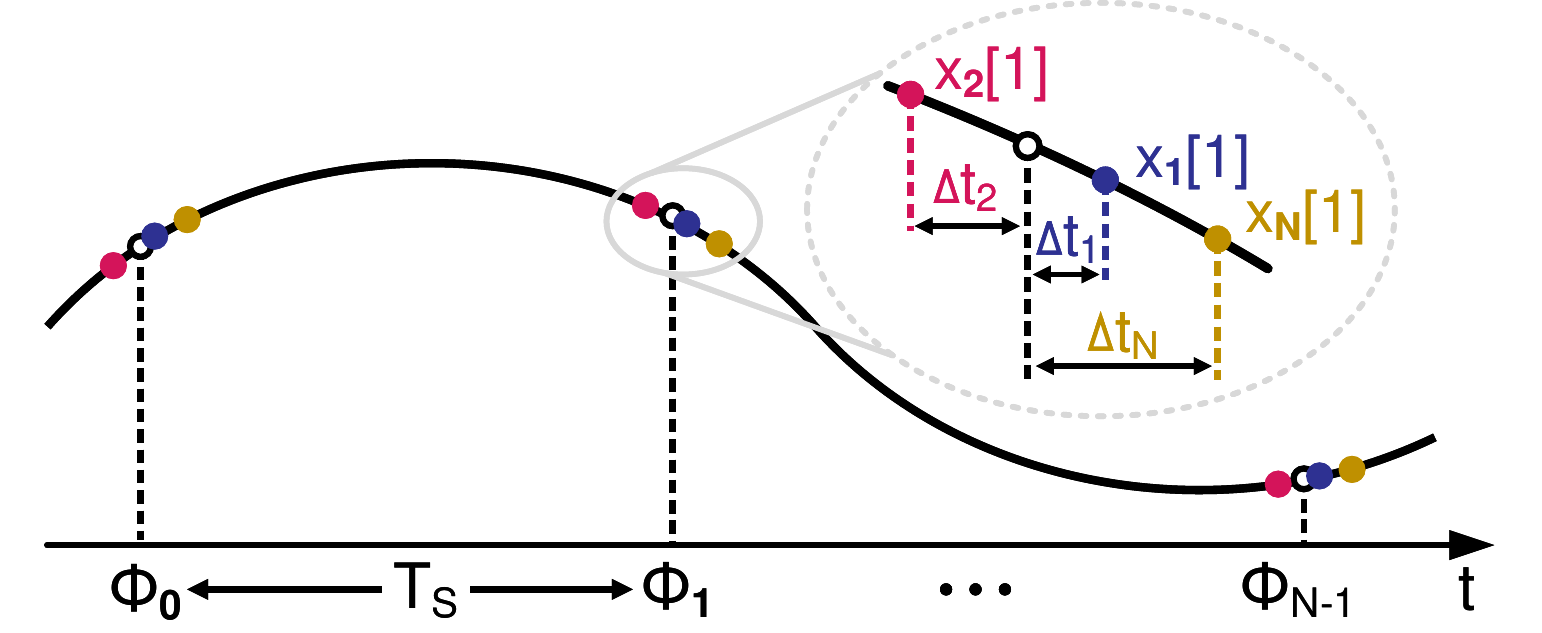}
	} 
	\hfill
	\subfloat[] 
	{ 
		\label{fig-2.3b} 
		\includegraphics[width = 0.48\linewidth]{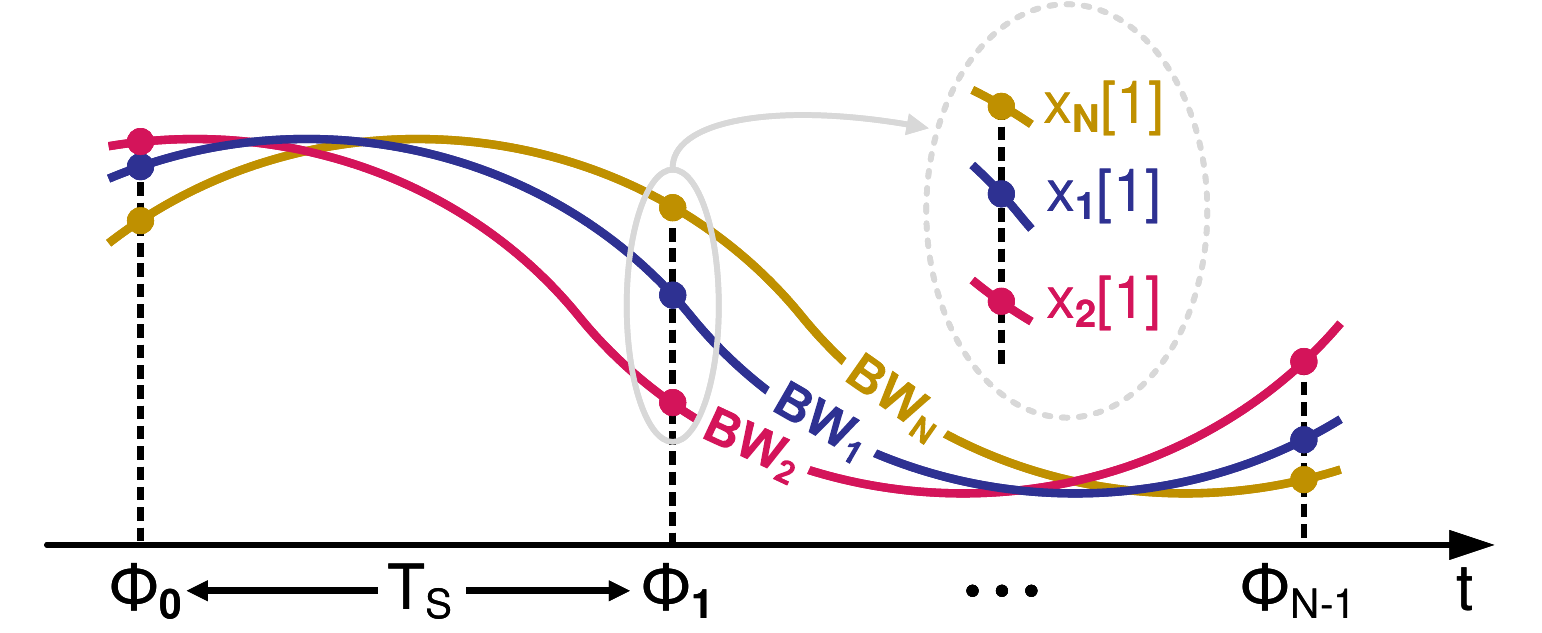} 
	}
	\caption{\revision{Illustration of timing mismatch errors introduced by (a) timing skew and (b) tracking bandwidth mismatch.}} 
	\label{fig-2.3} 
\end{figure*}

\figurename\ref{fig-2.3} further illustrates the timing mismatch errors induced by timing skew and tracking bandwidth variations. 
The timing skew is equivalent to the phase modulation of the input signal, whereas the bandwidth mismatch is reflected in the frequency responses of the individual channels, resulting in both gain and phase mismatches.



\subsection{Timing Skew}\label{skew_analysis}

Assuming a single-tone sinusoidal input, and given the timing skews $\Delta t_{i}$ across the $N$ channels exhibit a periodic pattern with a period of $N/f_{\mathrm{s}}$ \cite{2016High}, the input to the $i$-th channel can be expressed as
\begin{align}\label{eq2.4}
	x_{i} \left ( t \right ) =&\sin \left [ 2\pi f_{\mathrm{in}}\left ( t-\Delta t_{i}  \right )   \right ]	\notag
	\\= & \sin \left ( 2\pi f_{\mathrm{in}} t \right ) \cos  \left ( 2\pi f_{\mathrm{in}} \Delta t_{i}  \right )
	\\&- \cos \left ( 2\pi f_{\mathrm{in}} t \right ) \sin  \left ( 2\pi f_{\mathrm{in}} \Delta t_{i}  \right )\notag.
\end{align}
For a small skew, (\ref{eq2.4}) can be approximated as
\begin{align}\label{eq2.5}
	x_{i} \left ( t \right ) \approx \sin \left ( 2\pi f_{\mathrm{in}} t \right ) - \left ( 2\pi f_{\mathrm{in}} \Delta t_{i} \right ) \cos  \left ( 2\pi f_{\mathrm{in}} t \right ). 
\end{align}
$x_{i} \left ( t \right ) $ is sampled to $x_{i} \left [ n \right ] $ and subsequently transformed into frequency-domain components,
\begin{align}\label{eq2.6}
	Y\left [ k \right ]= \sum_{n=0}^{N-1}x_{i} [n]e^{-j\frac{2\pi}{N}kn}, \quad i=1,2,...,N; k=i-1.
\end{align}
And the IDFT reconstructs the time-domain signal from $Y\left [ k \right ]$,
\begin{align}\label{eq2.61}
	\mathsf{x'}\left [ m \right ]=\frac{1}{N}\sum_{k=0}^{N-1}Y\left [ k \right ]e^{j\frac{2\pi }{N}km } , \quad m=0,1,...,N-1.
\end{align}
From (\ref{eq2.5})-(\ref{eq2.61}), the reconstructed time-domain signal can be derived as
\begin{equation}\label{eq2.7}
	\begin{aligned}
		\mathsf{x'}\left(\frac{m}{f_\mathrm{s}}\right)
		=&\;
		\sin\left(2\pi f_{\mathrm{in}}\frac{m}{f_\mathrm{s}}\right) \\
		&-\frac{2\pi f_{\mathrm{in}}}{N}
		\sum_{n=0}^{N-1}
		\Biggl[
		\Bigl(
		\sum_{i=1}^{N}
		e^{-\frac{j2\pi(n-m)(i-1)}{N}}
		\Delta t_i
		\Bigr)
		\\
		&\quad\cdot
		\cos\Bigl(2\pi f_{\mathrm{in}}\frac{n}{f_\mathrm{s}}\Bigr)
		\Biggr],
		\quad
		m=0,\ldots,N-1
	\end{aligned}
\end{equation}
where the first term is the ideal output and the second term represents the error caused by timing skews. The power of the skew error can be expressed as
\begin{equation}\label{eq2.8}
	\begin{aligned}
		P_{\mathrm{skew,AFT}}
		=&\;
		\frac{(2\pi f_{\mathrm{in}})^{2}\sin^{2}(N\pi f_{\mathrm{in}}/f_\mathrm{s})}
		{2N^{3}\sin^{2}(\pi f_{\mathrm{in}}/f_\mathrm{s})}
		\sum_{i=1}^{N}\Delta t_i^{2}
		\\
		=&\;
		\frac{\sin^{2}(N\pi f_{\mathrm{in}}/f_\mathrm{s})}
		{N^{2}\sin^{2}(\pi f_{\mathrm{in}}/f_\mathrm{s})}
		\cdot P_{\mathrm{skew,TI}}
		\\
		<&\;
		P_{\mathrm{skew,TI}}
	\end{aligned}
\end{equation}
where $P_\mathrm{skew,TI}$ represents the skew error power in a TI system \cite{2016High}. 
Owing to the periodic sinc function, the power of the skew error in the proposed AFT-based FI system is significantly attenuated compared to its TI counterpart.
In particular, the system is theoretically immune to timing skews when the input frequency $f_{\mathrm{in}}$ is an integer multiple of $f_\mathrm{s}/N$.

\begin{figure} [!t]
	\centering 
	\subfloat[$\sigma_{\mathrm{skew}} =2\%$]	
	{
		\label{fig-2.4a} 
		\includegraphics[width = 0.97\linewidth]{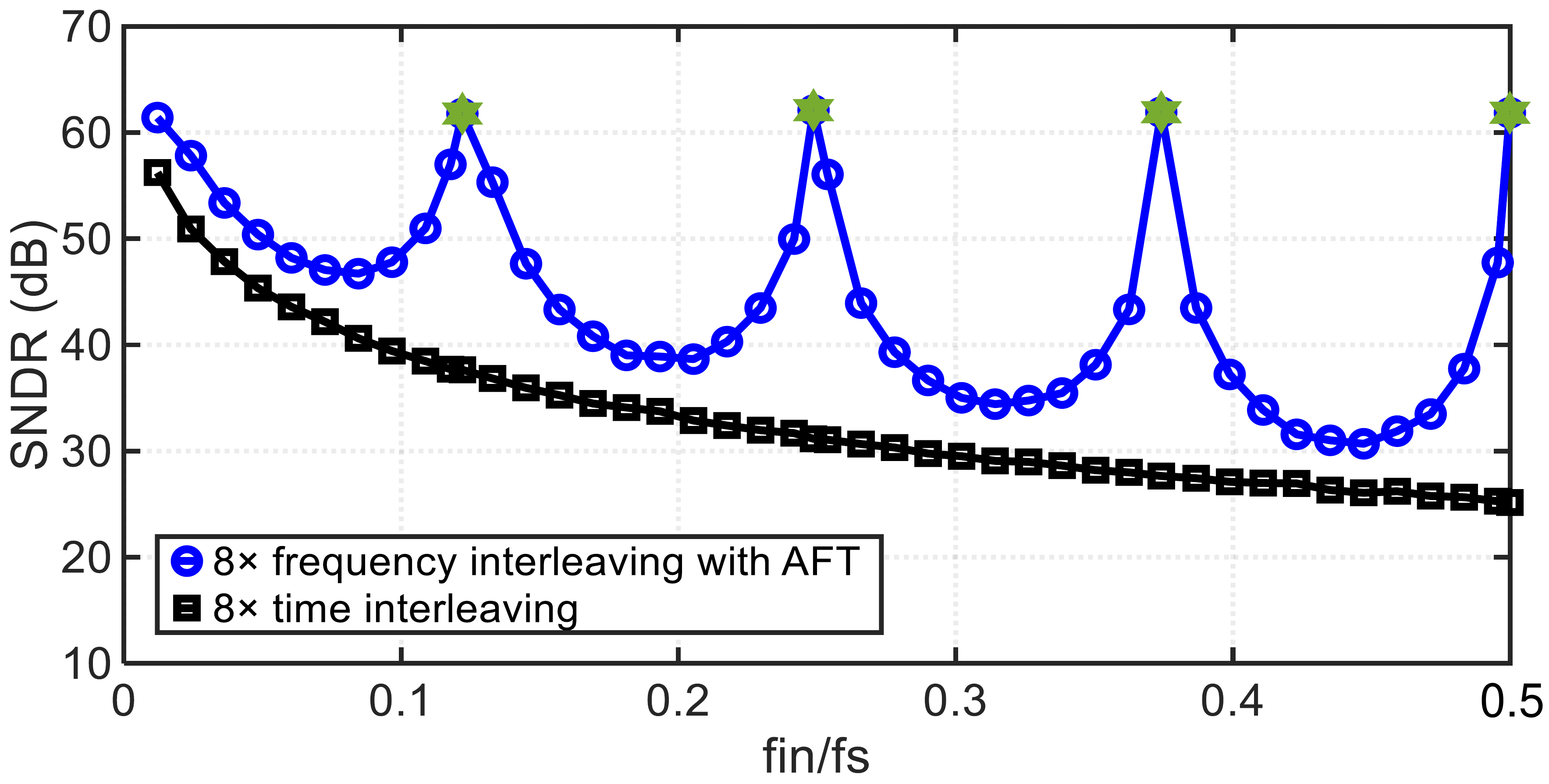}
	} 
	\\
	\subfloat[$\sigma_{\mathrm{RC}} =2\%$]	 
	{
		\label{fig-2.4b} 
		\includegraphics[width = 0.97\linewidth]{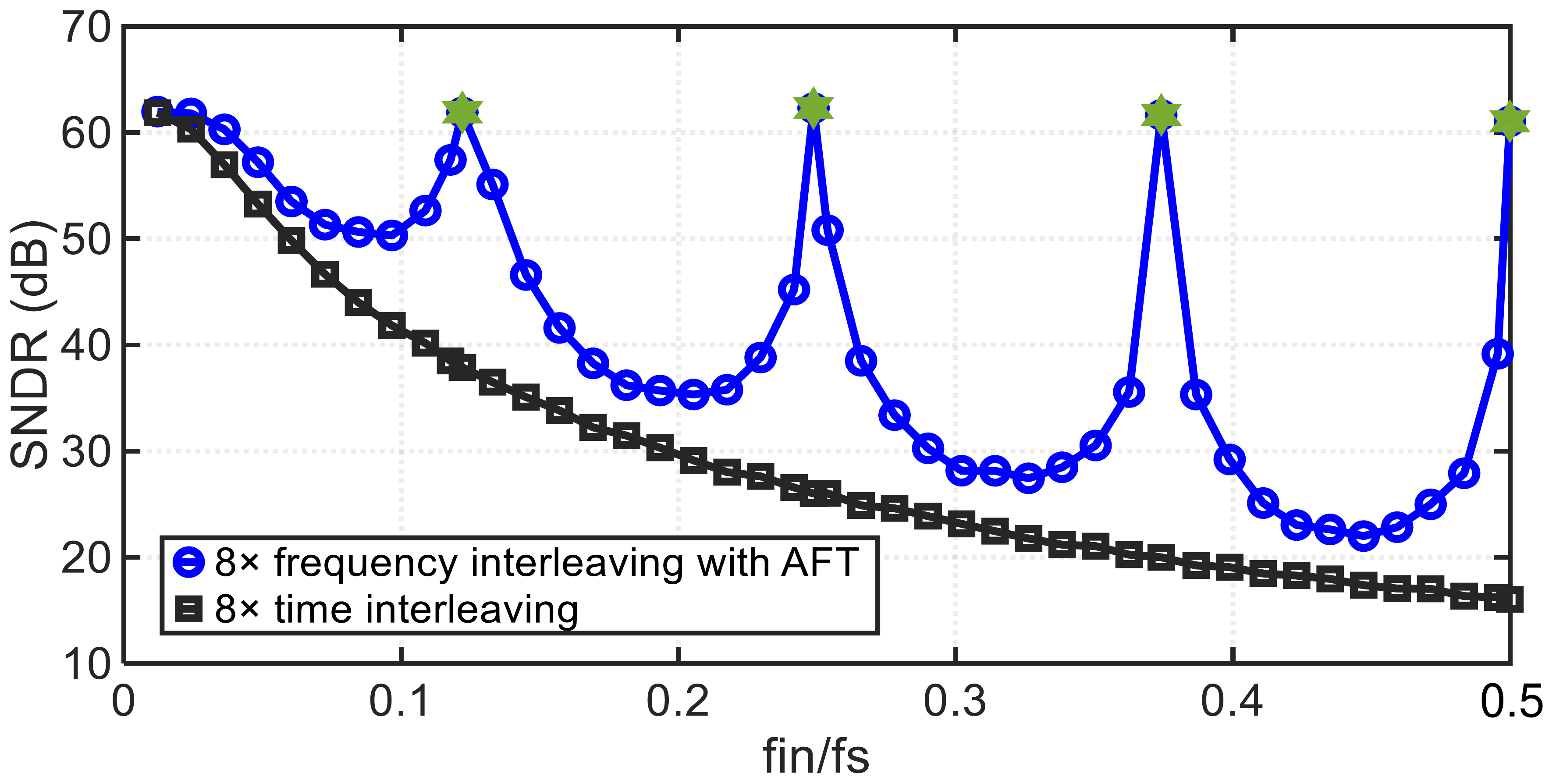}
	} 
	\caption{Behavioral simulation of SNDR versus normalized input frequency $f_\mathrm{in}/f_\mathrm{s}$ under (a) timing skew (b) tracking bandwidth mismatch for an 8-point AFT-based FI ADC and an 8-channel TI ADC, both with ideal 10b sub-ADCs.}
	\label{fig-2.4}
\end{figure}

Behavioral-level simulations of the proposed FI system have been performed, where an 8-point AFT-based FI ADC with ideal 10b sub-ADCs is compared against an 8-channel TI ADC using identical 10b sub-ADCs.
\figurename\ref{fig-2.4}\subref{fig-2.4a} shows the simulated SNDR versus the normalized input frequency $f_\mathrm{in}/f_\mathrm{s}$ under a timing skew with a standard deviation of 2\% of the sampling period ($\sigma_\mathrm{skew}=2\% T_\mathrm{s}$).
The proposed FI system remains error-free at $f_\mathrm{in}/f_\mathrm{s}$ of 1/8, 1/4, 3/8 and 1/2, while achieving 5dB improvement in the worst case compared to the conventional TI system.

\begin{figure} []
	\centering 
	\subfloat[] 
	{ 
		\label{fig-2.5a} 
		\includegraphics[width = 0.90\linewidth]{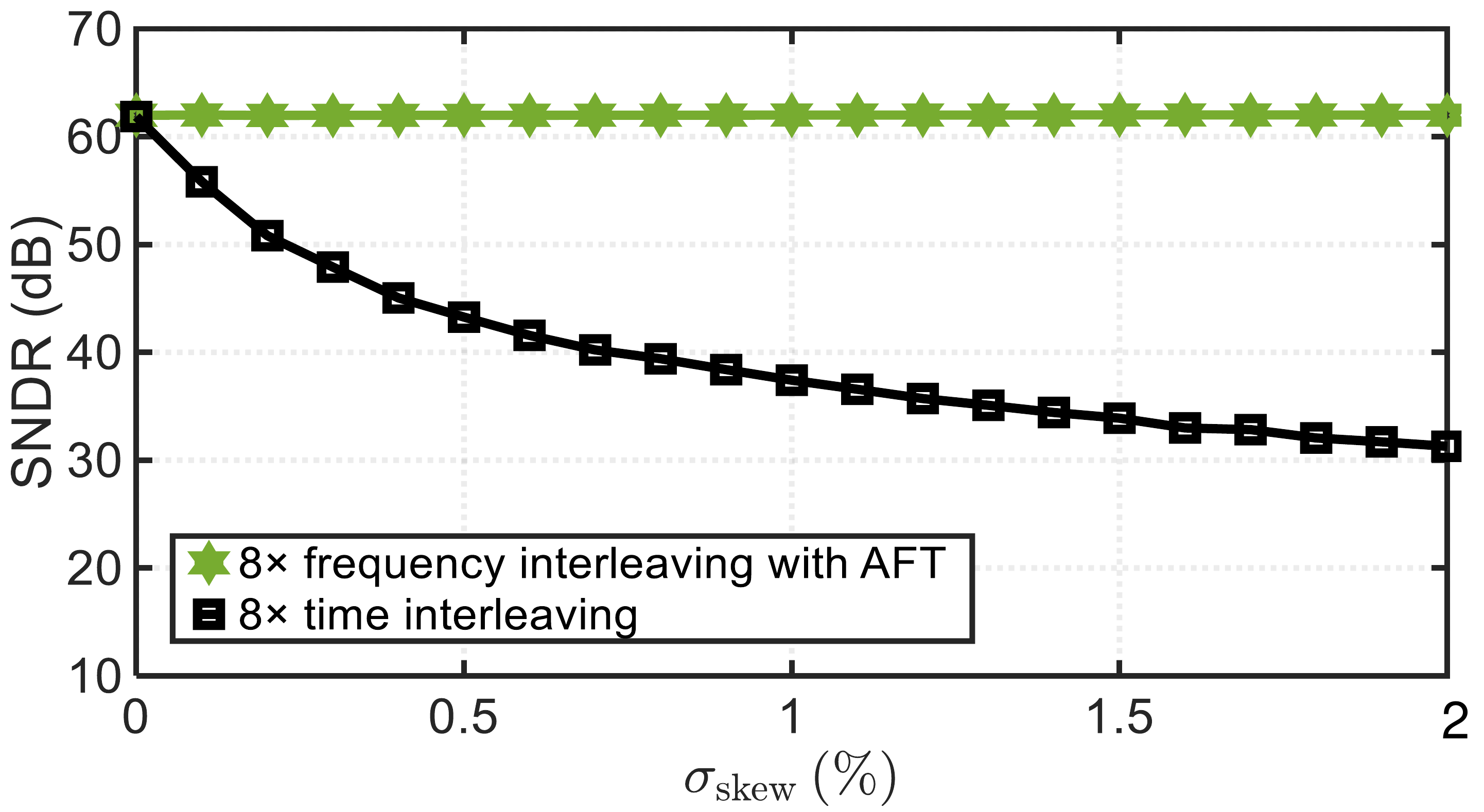} 
	}
	\\
	\subfloat[] 
	{ 
		\label{fig-2.5b} 
		\includegraphics[width = 0.90\linewidth]{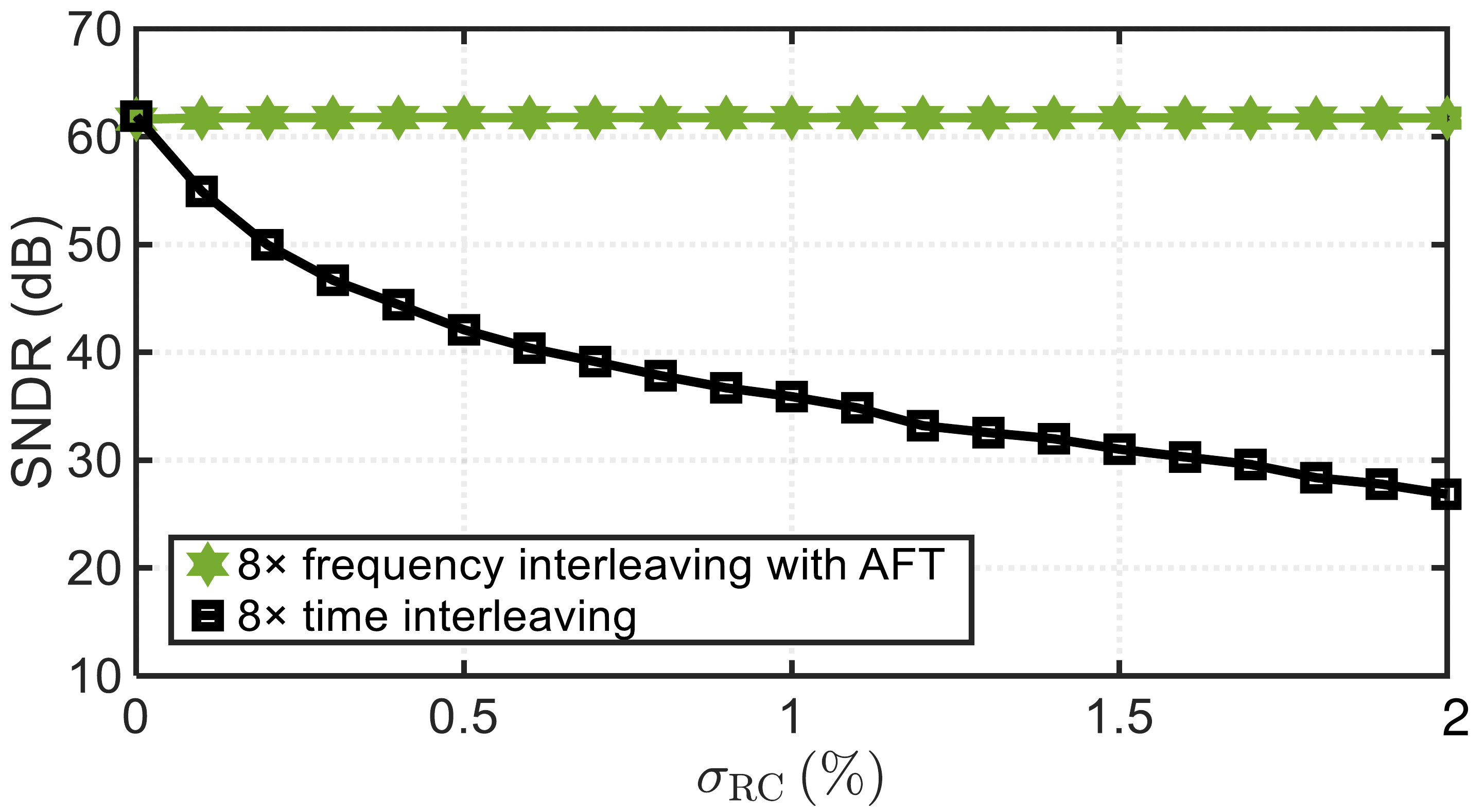} 
	}
	\caption{Behavioral simulation of SNDR versus (a) timing skew and (b) bandwidth mismatch. The input frequency $f_\mathrm{in}$ is set as $1/4 f_\mathrm{s}$.}
	\label{fig-2.5}
\end{figure}


\subsection{Tracking Bandwidth Mismatch}\label{bwm_analysis}

With tracking bandwidth mismatch, the input signal of the $i$-th channel can be written as
\begin{equation}\label{eq2.9}
	x_{i}\left ( t  \right ) =\left ( 1-\Delta G_{i}  \right ) \sin \left ( 2\pi f_{\mathrm{in}} t - \Delta \theta _{i}  \right )
\end{equation}
where $\Delta G_{i}$ and $\Delta \theta_{i}$ represent the gain and phase mismatches caused by bandwidth mismatch \cite{2016High}.
Considering a simplified first-order $RC$ network, the bandwidth-mismatch error power in the AFT-based FI system can be derived through a similar procedure as
\begin{equation}\label{eq2.10}
	P_\mathrm{BM,AFT} \approx  \frac{\sin^{2}  \left (N\pi f_{\mathrm{in}}/f_{\mathrm{s}}\right ) }{N^{2}\sin ^{2}\left ( \pi f_{\mathrm{in}}/f_{\mathrm{s}}   \right )   }\cdot P_\mathrm{BM,TI} < P_\mathrm{BM,TI}
\end{equation}
Consequently, the AFT-based FI system inherits the same attenuation factor determined by the periodic sinc function, effectively suppressing the bandwidth-mismatch errors compared to its TI counterpart.

\figurename\ref{fig-2.4}\subref{fig-2.4b} shows the simulated SNDR versus the normalized input frequency under a bandwidth mismatch modeled by a first-order $RC$ network with $\sigma_{\mathrm{RC}} =2\%$.
The results exhibit a performance trend highly consistent with that observed during the skew analysis, validating the inherent robustness of the AFT-based FI system against tracking bandwidth mismatches.

\subsection{System Perspective}\label{aft_system}

It is worth further discussing the specific frequency points where the proposed AFT-based FI architecture is theoretically immune to timing errors.
As mathematically proven in (\ref{eq2.8}) and (\ref{eq2.10}), when the input frequency $f_{\mathrm{in}}$ satisfies $f_{\mathrm{in}} = n \cdot f_{\mathrm{s}}/N$ ($n \in \mathbb{Z}$), the periodic sinc factor drops to zero, effectively creating a notch-filtering effect on the timing errors.

\figurename\ref{fig-2.5} shows the simulated SNDR versus skew and bandwidth mismatch when the input frequency $f_\mathrm{in}$ is set at $1/4 f_\mathrm{s}$. 
The proposed AFT-based FI architecture exhibits a prominent advantage over the conventional TI architecture at these specific synchronous frequencies, particularly as the timing errors exacerbate.

Admittedly, this condition cannot be generally fulfilled in blind ADC applications, such as real-time oscilloscopes and wideband spectrum analyzers, where the input signal is arbitrary and unknown a priori.
However, this unique characteristic is highly practical and can be readily leveraged in various modern systems.  
For instance, in coherent sampling, the input frequency is constrained to be a rational multiple of $f_\mathrm{s}/N$.
Similarly, in OFDM-based systems, the sub-carrier frequencies are naturally aligned with integer multiples of the spacing ($f_\mathrm{s}/N$).
Consequently, the timing-error immunity of the proposed AFT-based FI architecture can eliminate the need for power-hungry calibrations, manifesting a significant advantage from the system perspective.

%% file: sections/3_implementation_of_the_irrational_twiddle_factor.tex
\section{Circuit Implementation of Complex-valued Irrational Coefficients}\label{sec:implementation_of_the_irrational_twiddle_factor}

\begin{figure}[t]
	\centering
	\includegraphics[width = 0.95\linewidth]{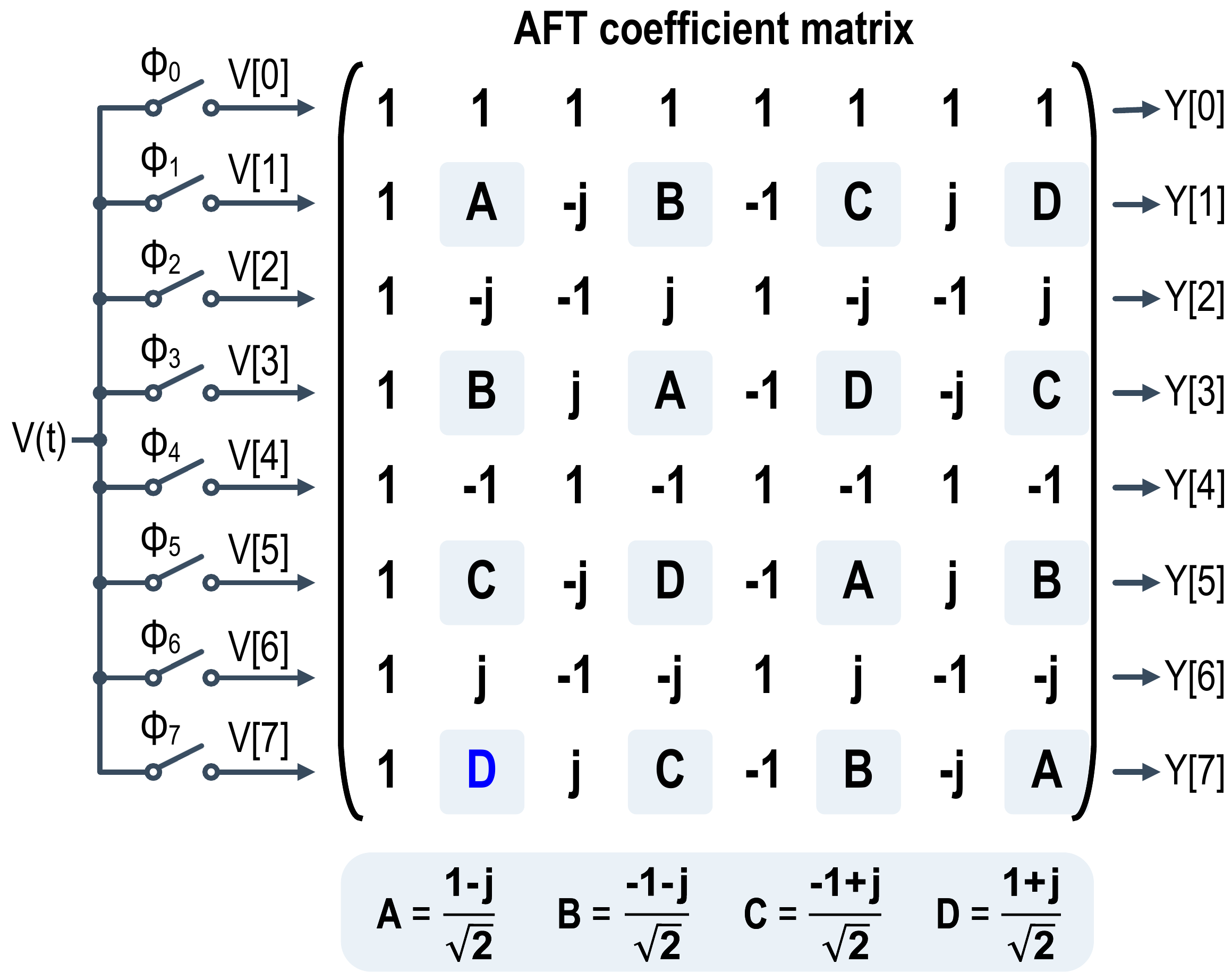}
	\caption{Coefficient matrix for an 8-point AFT.}
	\label{fig-3.1}
\end{figure}
\begin{figure} [t]
	\centering 
	\subfloat[]	
	{
		\label{fig-3.2a} 
		\includegraphics[width = 0.466\linewidth]{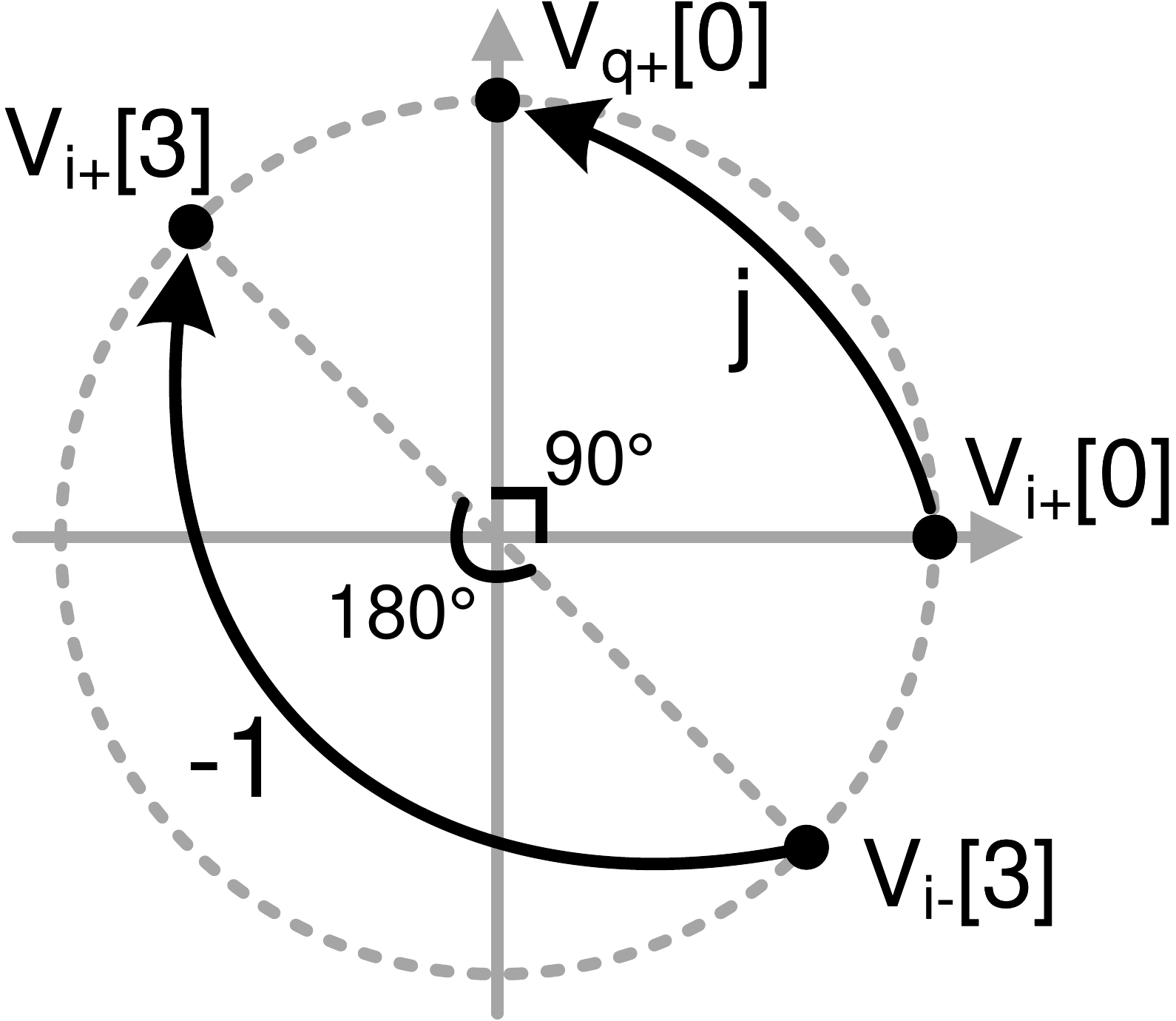}
	} 
	\hfill
	\subfloat[] 
	{ 
		\label{fig-3.2b} 
		\includegraphics[width = 0.466\linewidth]{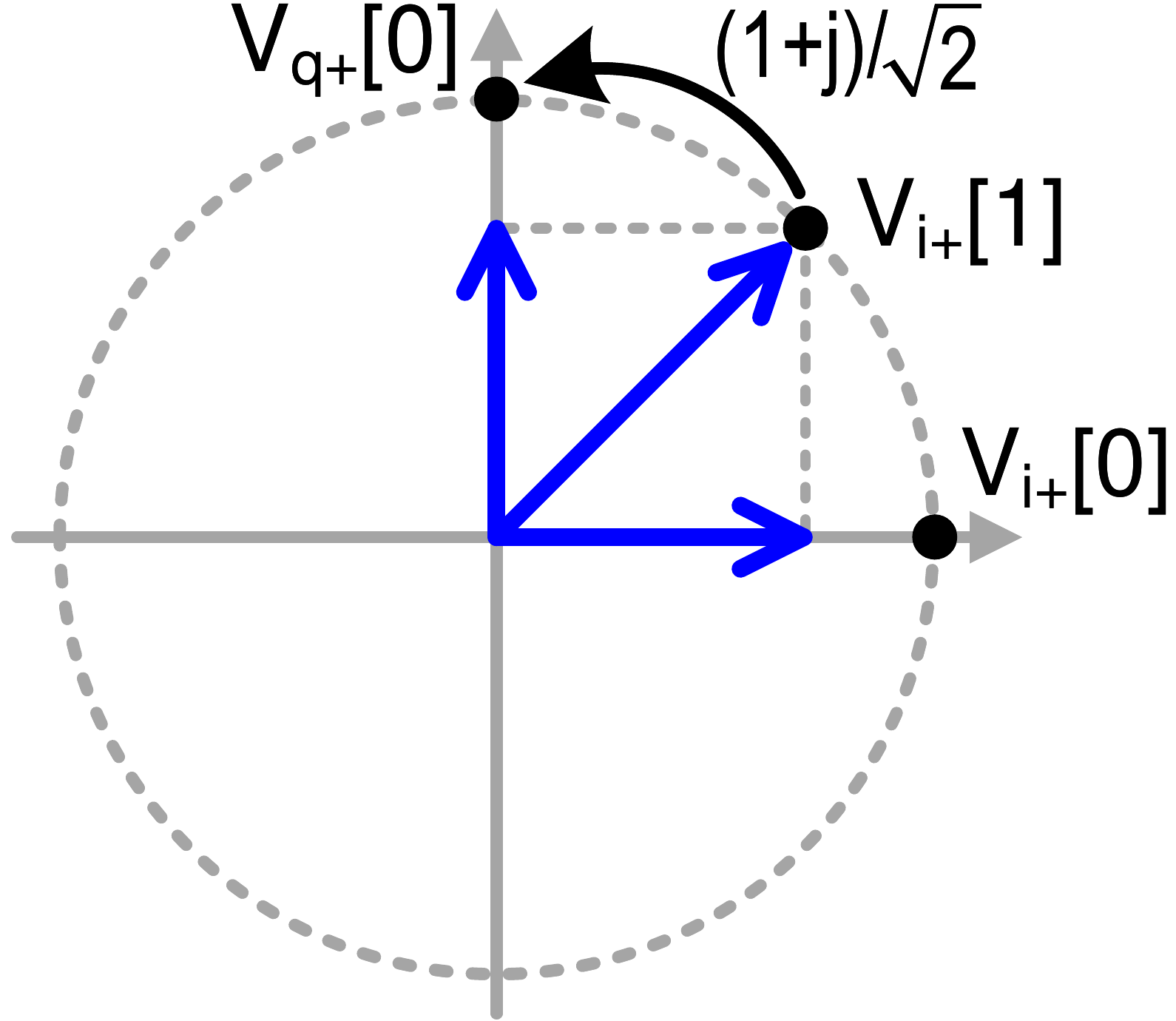} 
	}
	\caption{Unit-circle illustration of the (a) rational and (b) irrational complex-valued twiddle factors.}
	\label{fig-3.2} 
\end{figure}

\begin{figure} [t]
	\centering 
	\subfloat[sampling]	
	{
		\label{fig-3.4a} 
		\includegraphics[width = 0.99\linewidth]{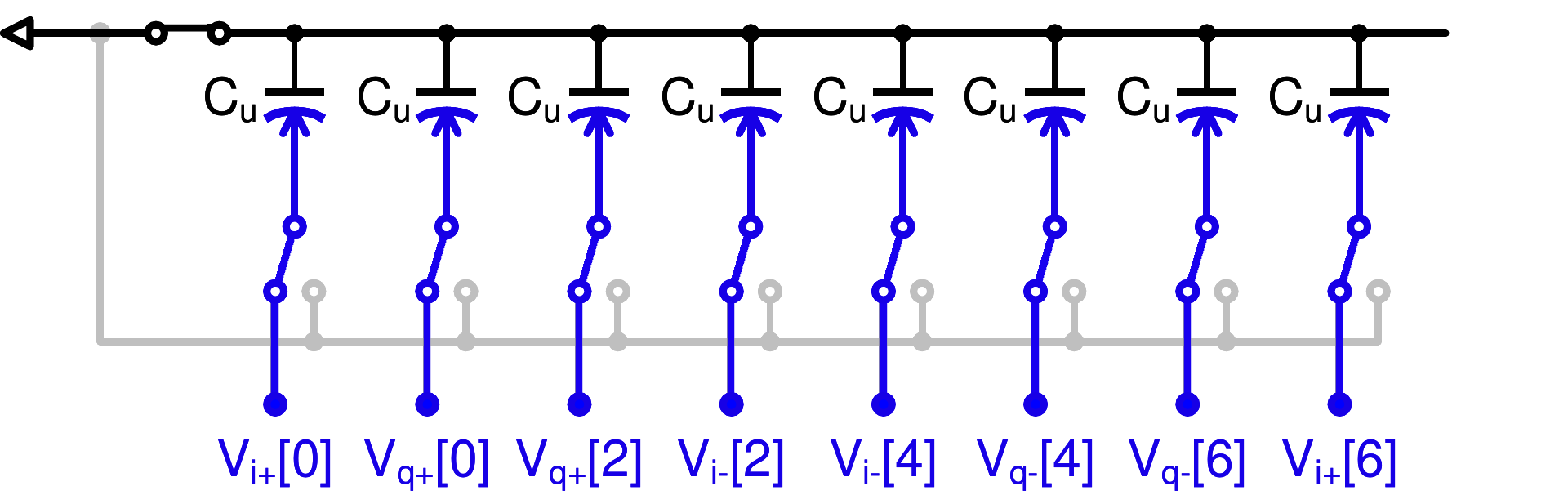}
	} 
	\newline
	\subfloat[AFT processing] 
	{ 
		\label{fig-3.4b} 
		\includegraphics[width = 0.99\linewidth]{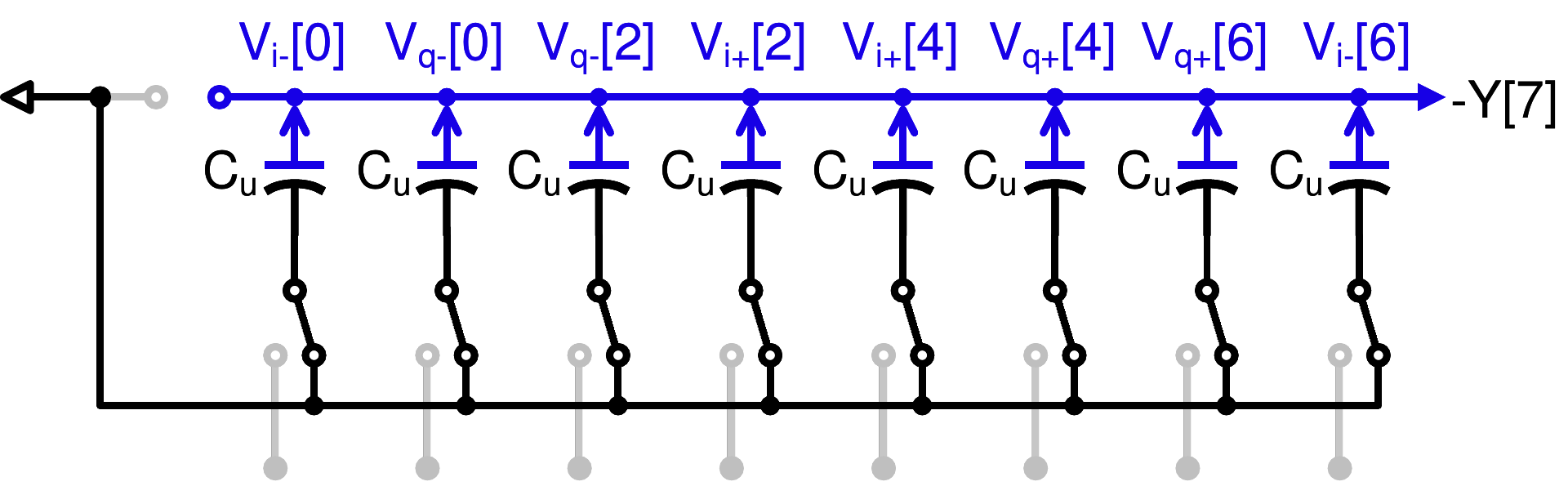} 
	}
	\caption{Single-ended AFT implementation example for the $Y[7]$ channel.} 
	\label{fig-3.4} 
\end{figure}


As indicated by (\ref{eq2.8}) and (\ref{eq2.10}), a larger AFT point $N$ (i.e., a higher number of FI channels) provides a higher level of robustness against timing errors.
The trade-off here is the implementation complexity of the AFT, where a large complex coefficient matrix with irrational numbers has to be realized.
Ratio-based implementations of the twiddle factors have been exploited in the current domain \cite{TCAS2010-Lechne} and the charge domain \cite{JSSC2013-Sadhu,CICC2015-Shin}. 
However, there are inevitable errors for the irrational numbers, resulting in systematic errors that limit the linearity of the AFT calculation.
In this work, a lossless implementation of an 8-point AFT is proposed.

\figurename\ref{fig-3.1} shows the coefficient matrix for an 8-point AFT, where the coefficients $A$, $B$, $C$ and $D$ are complex-valued irrational numbers containing $\sqrt{2}$.
The input signal $V\left ( t \right ) $ is sampled into $V\left [ 0 \right ]$--$V\left [ 7 \right ]$ at phases $\phi_{0}$--$\phi_{7}$, and $Y\left [ 0 \right ]$--$Y\left [ 7 \right ]$ are the frequency-domain outputs of the AFT.

With the help of differential I/Q signaling, the rational coefficients can be implemented conveniently.
As depicted in \figurename\ref{fig-3.2}\subref{fig-3.2a}, the complex operation $V_{i+}\left [ 0 \right ] \times j$ is equivalent to a $90^{\circ }$ counterclockwise phase shift of $V_{i+}\left [ 0 \right ]$, corresponding to $V_{q+}\left [ 0 \right ]$.
Similarly, $V_{i+} \left [ 3 \right ] \times \left ( - 1 \right )$ equals to a $180^{\circ }$ phase shift of $V_{i+} \left [ 3 \right ]$, corresponding to $V_{i-}\left [ 3 \right ]$.

\figurename\ref{fig-3.2}\subref{fig-3.2b} illustrates the calculation with an irrational coefficient, for instance $V_{i+} \left [ 1  \right ] \times \left (1+j \right )/ \sqrt{2}$, as the second term of $Y[7]$.
$V_{i+} \left [ 1  \right ]$ can be decomposed into two orthogonally sampled components
\begin{align}\label{eq_3.1}
V_{i+}\left [ 1  \right ] = \frac{V_{i+}\left [ 0 \right ]+  j V_{i+}\left [ 0  \right ]    }{\sqrt{2} }=\frac{V_{i+}\left [ 0  \right ]+  V_{q+}\left [ 0  \right ]    }{\sqrt{2} } ,
\end{align}
where an irrational scaling factor of $1/\sqrt{2}$ is generated. 
The derivation can then be evolved further
\begin{align}\label{eq_3.2}
V_{i+} \left [ 1  \right ] \times\frac{1+j}{\sqrt{2} } &=\frac{\left ( V_{i+}\left [0 \right ]+  V_{q+}\left [ 0 \right ] \right )\left ( 1+j \right )    }  {2} \notag \\
&= \frac{V_{i+}\left [0 \right ]+  jV_{i+}\left [ 0 \right ]+V_{q+}\left [0 \right ]+  jV_{q+}\left [ 0 \right ]}{2} \notag \\
&=\frac{V_{i+}\left [0 \right ]+  2V_{q+}\left [ 0 \right ]+  V_{i-}\left [ 0 \right ]}{2}  \notag \\
&= V_{q+} \left [ 0  \right ].
\end{align}
Therefore, the calculation of $V_{i+}[1]\times D$, a complex-valued irrational coefficient, can be simply implemented by re-wiring to $V_{q+}[0]$.
The calculations with the remaining irrational coefficients can be implemented following the same procedure.

\figurename\ref{fig-3.4} shows a single-ended example for the implementation of the $Y[7]$ channel, where all capacitors are with an equal size $\mathrm{C}_\mathrm{u}$.
During the sampling phase, the top plates are connected to ground while the corresponding I/Q signals are sampled onto $\mathrm{C}_\mathrm{u}$.
Subsequently, the top plate disconnects from ground and the bottom plates switch to ground simultaneously.
The AFT calculation is performed by charge redistribution.
Since the irrational calculation does not rely on device ratios, the AFT is free from systematic errors.


It is worth noting that, since the twiddle factors in DFT corresponds inherently to the complex exponentials on the unit circle, the proposed implementation can be generalized to AFT architectures of arbitrary point sizes with complex-valued irrational coefficients.

%% file: sections/4_circuit_implementation.tex
\section{Design of the Proposed AFT-based Frequency-Interleaved ADC}\label{sec:circuit_implementation}

\begin{figure*}[!t]
	\centering
	\includegraphics[width = 0.87\linewidth]{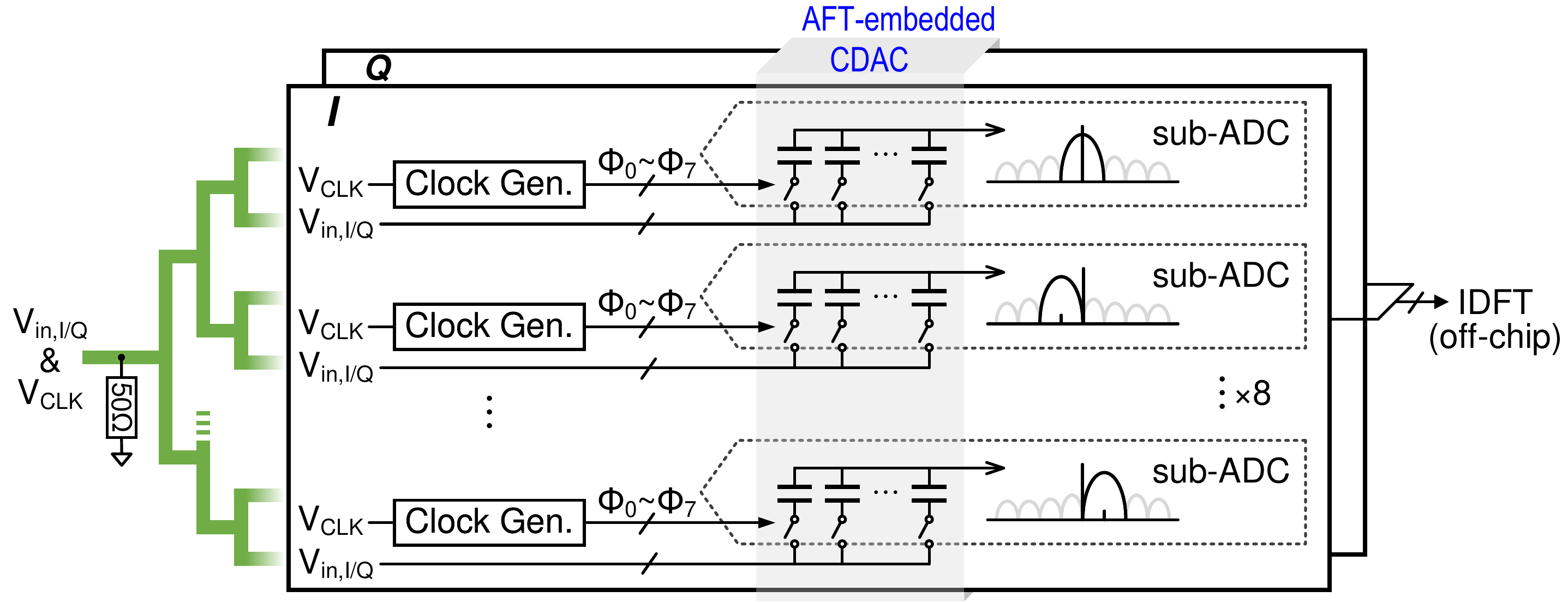}
	\caption{System architecture of the proposed 8-point AFT-based frequency-interleaved ADC.}
	\label{fig-4.1}
\end{figure*}


\figurename\ref{fig-4.1} shows the system architecture of the 8-point AFT-based FI ADC.
Each of the I/Q channels consists of eight frequency-interleaved sub-channels.
The input and clock signals are fed into each channel through a symmetric $Y$-tree-based distribution network.
Owing to the timing robustness offered by the proposed architecture, each channel generates its sampling clocks  $\phi _{0}$--$\phi _{7} $ locally, which significantly simplifies the design of the global clock distribution.
Furthermore, the AFT calculation is embedded into the sub-ADC through a partially shared capacitor array. 


\begin{figure*} [!t]
	\centering 

	{
		\label{fig-4.2a} 
		\includegraphics[width = 0.927\linewidth]{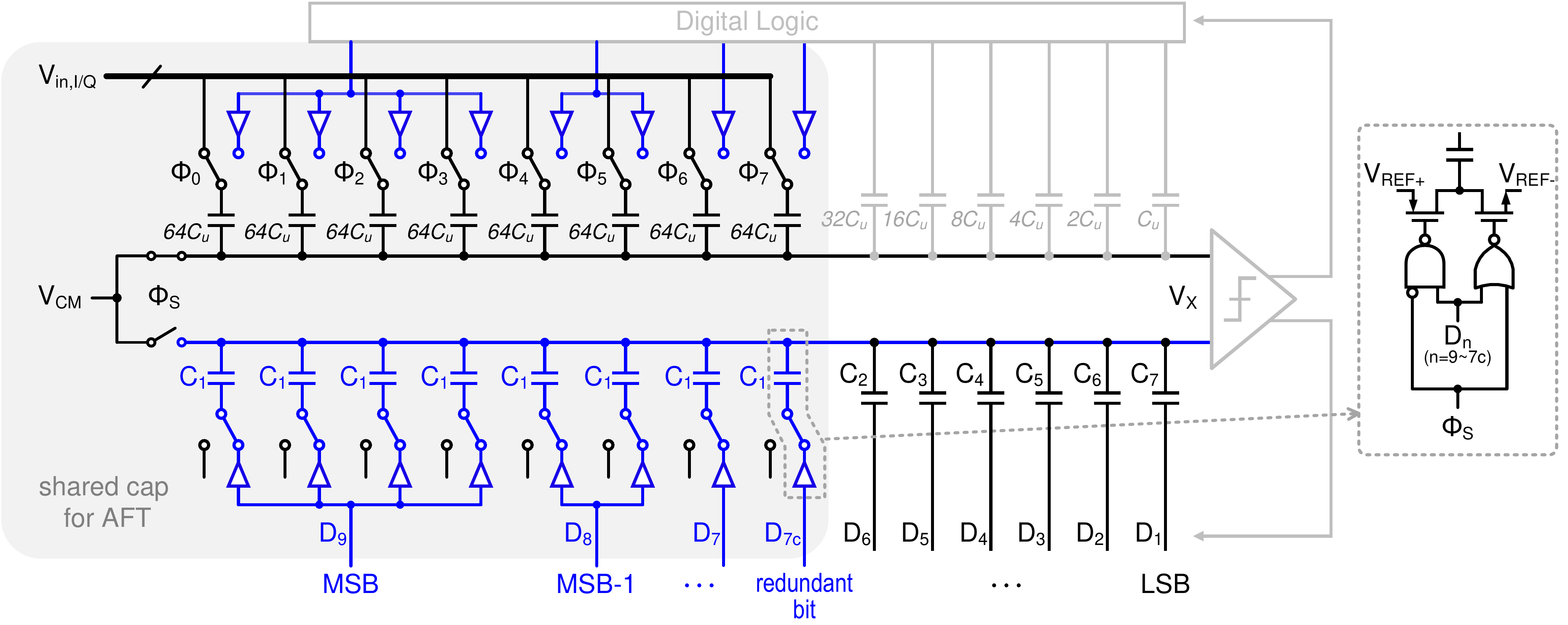}
	} 
	\caption{Single-channel implementation of the 10b SAR ADC with embedded AFT.} 
	\label{fig-4.2} 
\end{figure*}

\begin{figure} [!t]
	\centering 
	{ 
		\label{fig-4.2c} 
		\includegraphics[width = 0.99\linewidth]{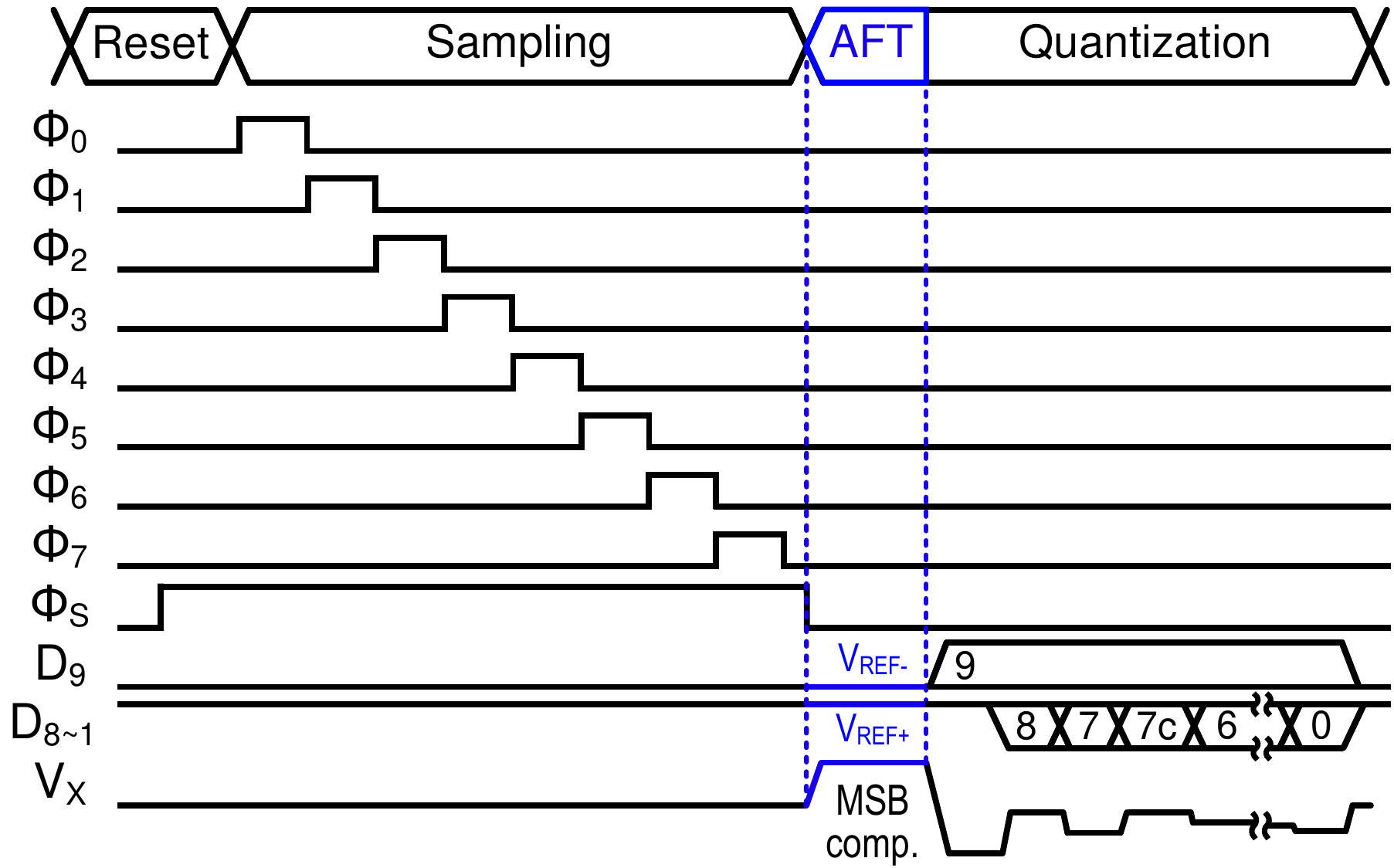} 
	}
	\caption{Timing diagram of the SAR ADC with embedded AFT.} 
	\label{fig-4.2c} 
\end{figure}

\subsection{AFT-Embedded CDAC}\label{subsec:Shared_CDAC_with_AFT_in_single-channel_ADC}

From \sectionname\ref{sec:implementation_of_the_irrational_twiddle_factor}, the charge-redistribution AFT can be embedded into the sampling process, and subsequently integrated into the CDAC of the sub-ADC. 
\figurename\ref{fig-4.2} shows the single-channel implementation of the 10b SAR ADC with embedded AFT, where the unit capacitance $\mathrm{C}_\mathrm{u}$ is 0.5 fF.
The timing diagram is depicted in \figurename\ref{fig-4.2c}, where $\phi_{0}$--$\phi_{7}$ are non-overlapping with a pulse width of 76.8 ps, resulting in a sampling rate of 13 GS/s.
The first 8 $\mathrm{C}_1$ are reused for both AFT and CDAC.

During the sampling phase, $\phi_\mathrm{S}$ is high, connecting the top plates to the common-mode voltage $\mathrm{V}_\mathrm{CM}$. 
The bottom plate of each $\mathrm{C}_{1}$ turns on at $\phi_{0}$--$\phi_{7}$ and samples the corresponding input depending on the respective coefficient.
The AFT calculation is performed subsequently, when $\phi_\mathrm{S}$ goes low, as described in \figurename\ref{fig-3.4}.

$\mathrm{C}_2$--$\mathrm{C}_7$ form a binary-scaled array with respect to $\mathrm{C}_1$, completing the full CDAC for the 10-bit SAR ADC, $[4\mathrm{C}_{1} , 2\mathrm{C}_{1}, \mathrm{C}_{1}, \mathrm{C}_{1}, \mathrm{C}_{2}, \mathrm{C}_{3}, \mathrm{C}_{4}, \mathrm{C}_{5}, \mathrm{C}_{6}, \mathrm{C}_{7}]$, where $\mathrm{C}_7$ corresponds to the unit capacitance $\mathrm{C}_\mathrm{u}$, and there is one redundant bit implemented by $\mathrm{C}_1$.
In this design, a monotonic switching scheme is exploited, and the MSB conversion can be performed concurrently with the AFT charge redistribution.
Since the MSB switching is normally the most time- and energy-consuming step, this combined AFT and MSB operation saves both conversion time and power.

Note that since $\mathrm{C}_2$--$\mathrm{C}_7$ are not utilized during the sampling phase, an attenuation factor of 0.89 is introduced, which is generally acceptable considering the embedded AFT implementation offers significant advantages in area, speed and power efficiency.




\begin{figure*}[t]
	\centering
	\includegraphics[width = 0.95\linewidth]{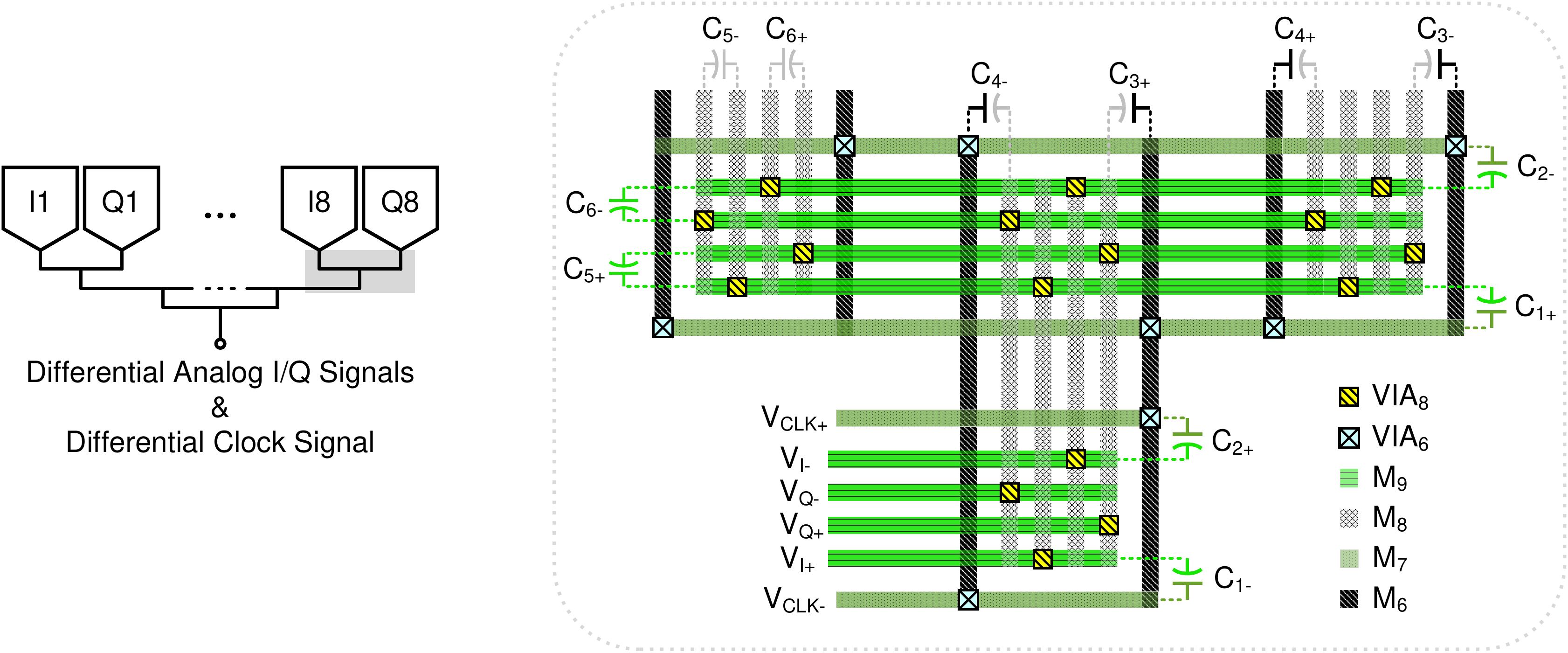}
	\caption{Layout illustration of the input and clock distribution network.}
	\label{fig-4.3}
\end{figure*}

\subsection{Input and Clock Distribution}\label{subsec:Passive_input_distribution_network}

For the proposed FI system, although the sensitivity to timing errors is significantly mitigated, maintaining the differential signaling and I/Q phase orthogonality remains critical.
In this design, a Y-tree network is exploited for the concurrent distribution of the differential clock and input I/Q signals, as illustrated in \figurename\ref{fig-4.3}, where the I/Q and clock traces are routed side-by-side to ensure matched lengths and associated parasitics. 

The differential routes for the input and clock signals are carefully swapped in each routing detour to achieve a symmetric parasitic environment.
Moreover, the cross-coupling effect between traces can be minimized \cite{JSSC2020-Ramkaj}.
As shown in \figurename\ref{fig-4.3}, there are three differential signals,  $\mathrm{V}_\mathrm{I}$, $\mathrm{V}_\mathrm{Q}$, and $\mathrm{V}_\mathrm{CLK}$, resulting in 6 traces and 12 cross-coupling capacitors between adjacent traces.
For instance, $\mathrm{C}_\mathrm{1+}$ is between $\mathrm{V}_\mathrm{I+}$ and $\mathrm{V}_\mathrm{CLK+}$, whereas $\mathrm{C}_\mathrm{1-}$ is between $\mathrm{V}_\mathrm{I+}$ and $\mathrm{V}_\mathrm{CLK-}$. 
$\mathrm{C}_\mathrm{1+}$ and $\mathrm{C}_\mathrm{1-}$ can be optimized at the layout level, e.g., through spacing adjustment, Via swapping, etc., such that they are matched.
Considering that $\mathrm{V}_\mathrm{CLK+}$ and $\mathrm{V}_\mathrm{CLK-}$ are differential, the overall cross-coupling effect on $\mathrm{V}_\mathrm{I+}$ is suppressed.

%% file: sections/5_measurement_results.tex
\section{Measurement Results}\label{sec:measurement_results}

\begin{figure}[t]
	\centering
	\includegraphics[width = 0.95\linewidth]{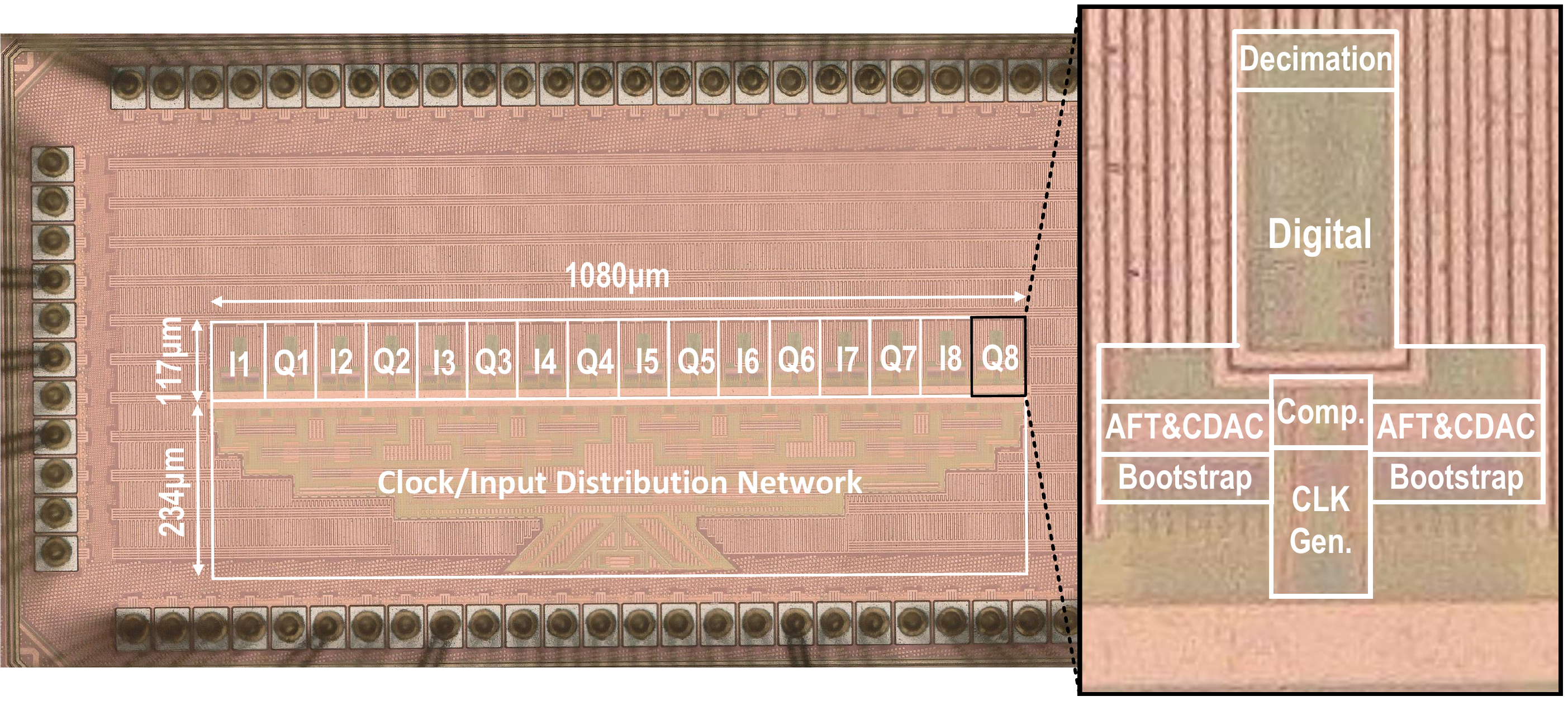}
	\caption{Die micrograph.}
	\label{fig-5.1}
\end{figure}
\begin{figure}[t]
	\centering
	\includegraphics[width = 0.99\linewidth]{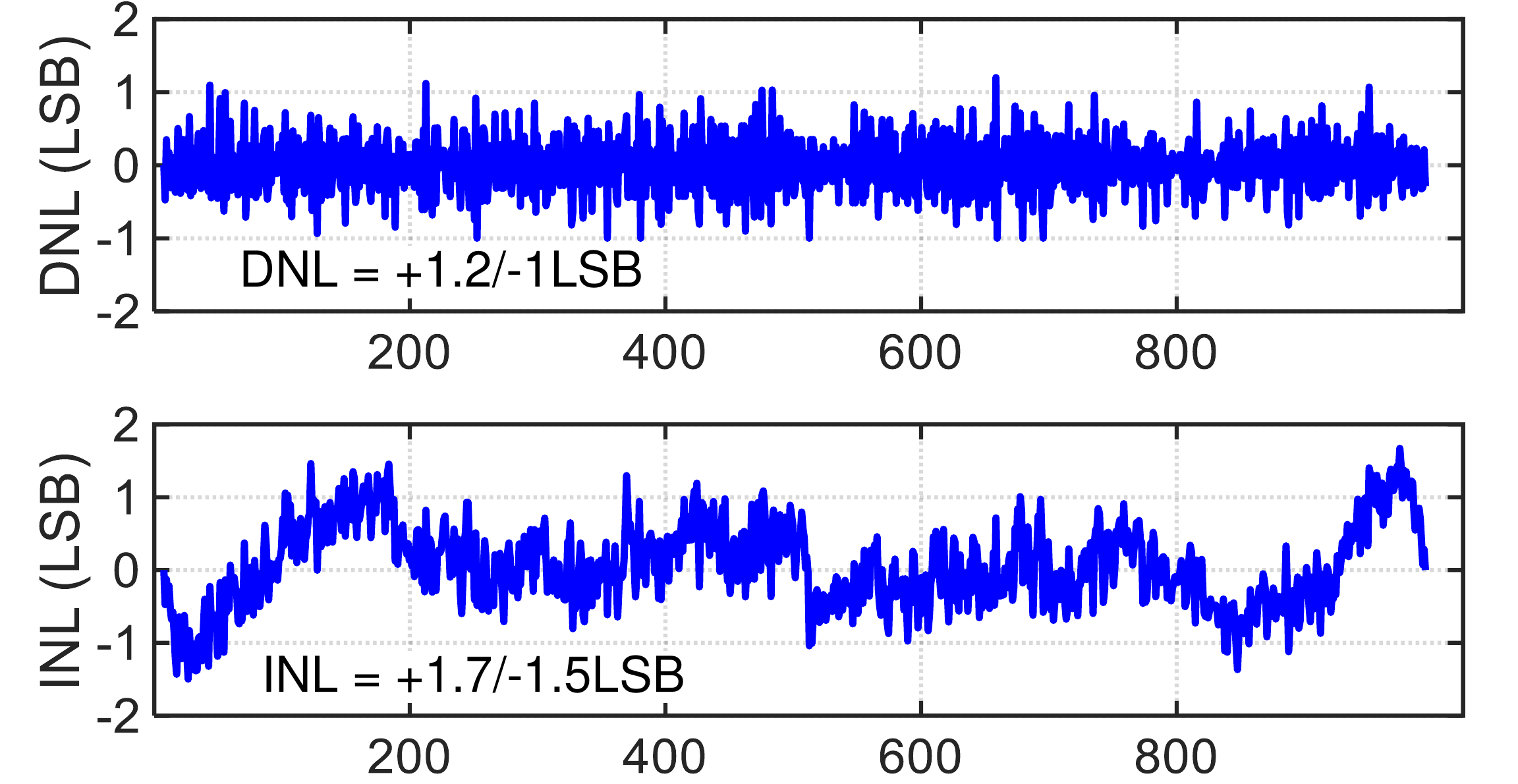}
	\caption{Measured static performance at 10b 13GS/s.}
	\label{fig-5.2}
\end{figure}

\begin{figure}[t]
      \subfloat[]	
	{
		\label{fig-5.3a} 
		\includegraphics[width = 0.93\linewidth]{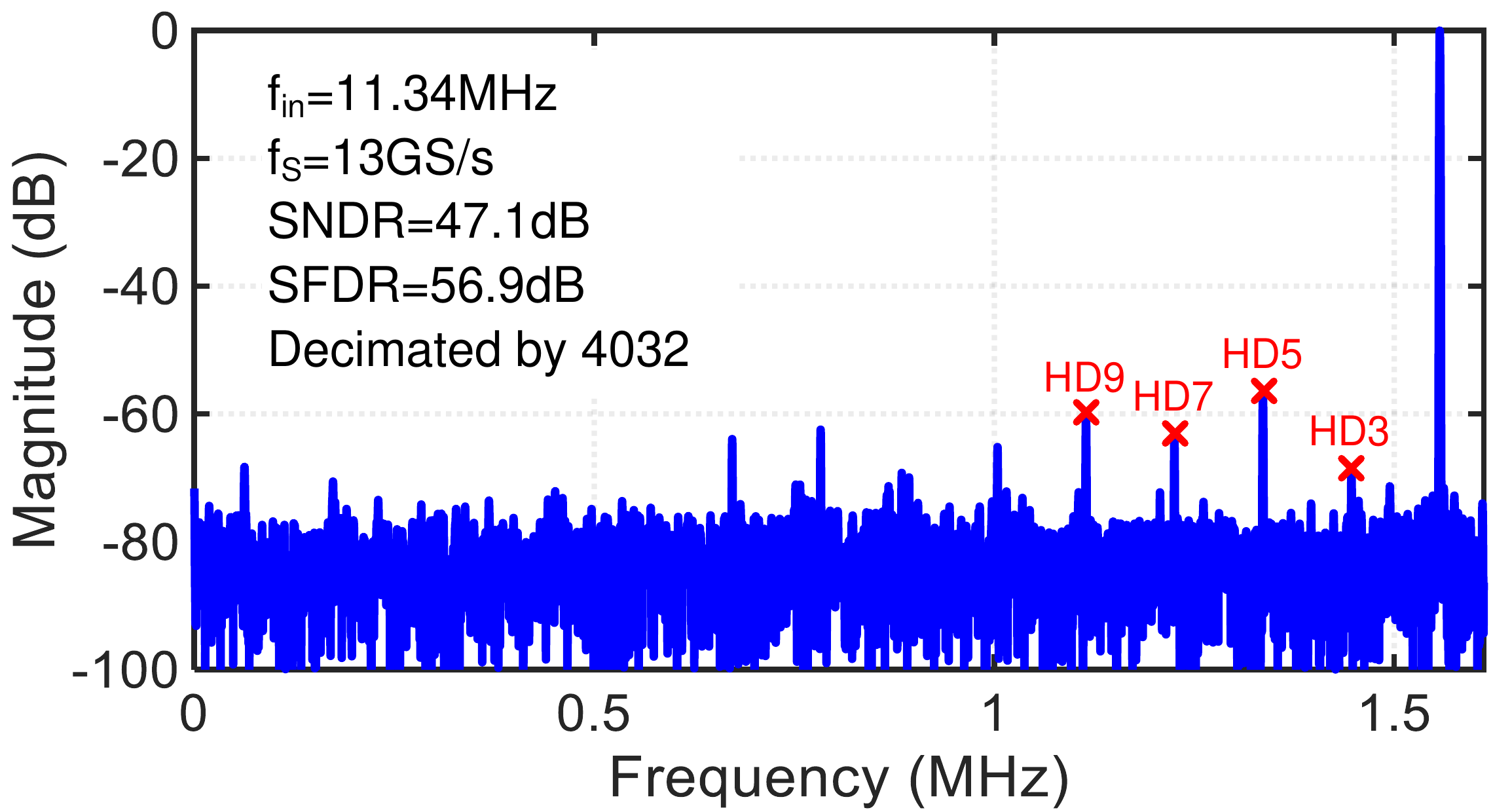}
	} 
	\newline
	\subfloat[] 
	{ 
		\label{fig-5.3b} 
		\includegraphics[width = 0.93\linewidth]{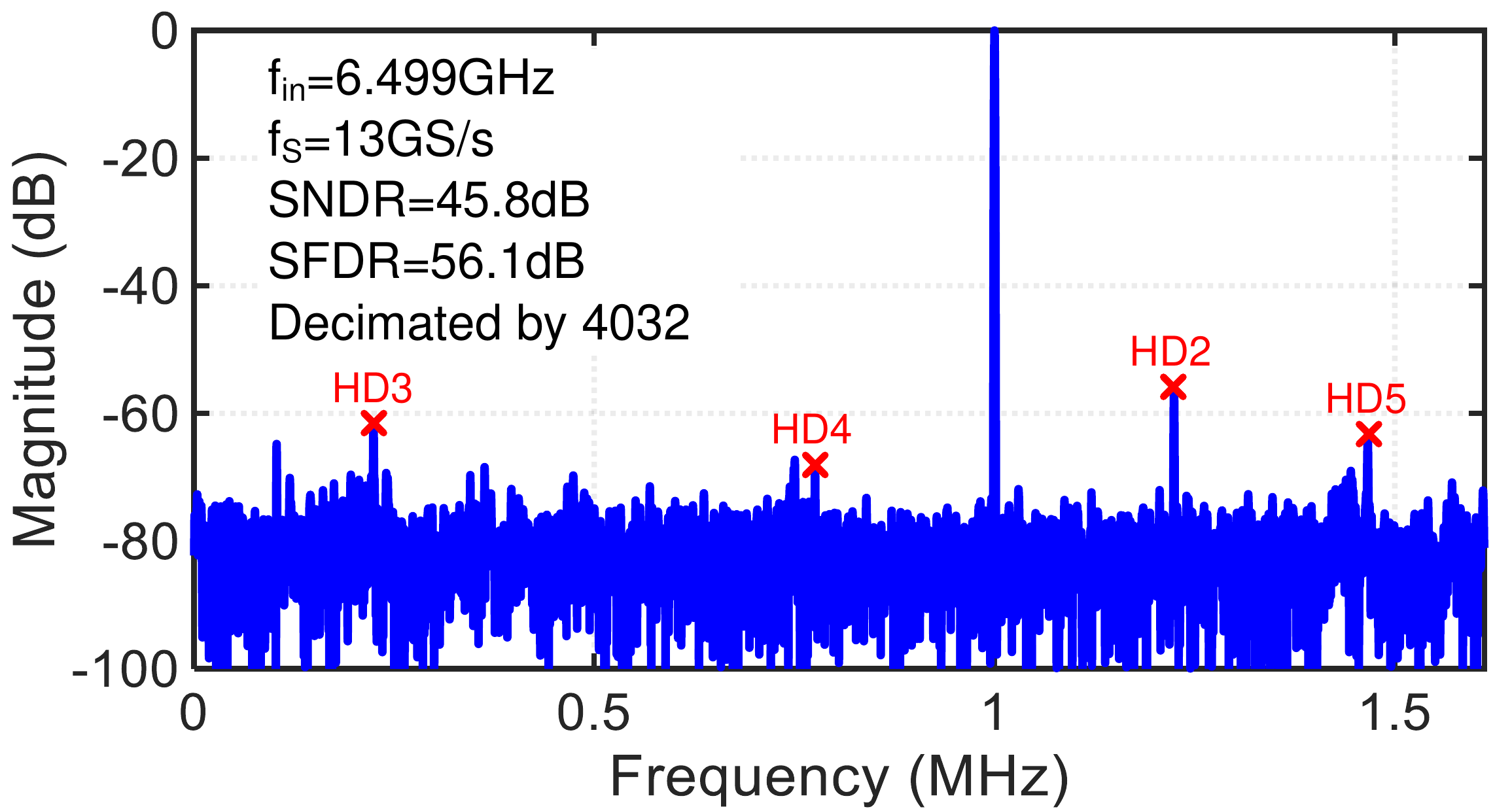} 
	}
	\caption{\revision{Measured spectra at (a) low frequency and (b) Nyquist.}}
	\label{fig-5.3}
\end{figure}

\begin{table*}[!t]
	\sffamily     
	\scriptsize						
	\renewcommand{\arraystretch}{1.7}	
	\newcommand{\tabincell}[2]{\begin{tabular}{@{}#1@{}}#2\end{tabular}}
	\centering
	\caption{Performance Summary and Comparison with Previous Published Works} \label{table_comparison}
	\begin{threeparttable}
		\begin{tabular}{c|c|cccccccc}
			\hline\hline\multicolumn{2}{c|}{Parameter}
				&\tabincell{c}{ISSCC 2017\\[-2mm]\cite{ISSCC2017-Keane}}
				&\tabincell{c}{JSSC 2021\\[-2mm]\cite{JSSC2021-Swindlehurst}}
				&\tabincell{c}{CICC 2024\\[-2mm]\cite{CICC2024-Zhang}}
				&\tabincell{c}{JSSC 2020\\[-2mm]\cite{JSSC2020-Zhang}}
				&\tabincell{c}{VLSI 2021\\[-2mm]\cite{VLSI2021-Zhang}}		
				&\tabincell{c}{ISSCC 2022\\[-2mm]\cite{ISSCC2022-Liu}}		                  
				&\tabincell{c}{VLSI 2024\\[-2mm]\cite{VLSI2024-Liu}}							
				&\textbf{This work}
			\\\hline\hline\multicolumn{2}{c|}{Architecture}
				&TI SAR
				&TI SAR
				&TI SAR
				&\tabincell{c}{Time-domain \\[-2mm] TI ADC}
				&\tabincell{c}{Time-domain \\[-2mm] TI ADC}
				&\tabincell{c}{Time-domain \\[-2mm] TI ADC}										
				&\tabincell{c}{Time-domain \\[-2mm] TI ADC}											
				&\textbf{\tabincell{c}{AFT-based \\[-2mm] FI ADC}}			
			\\\hline\multicolumn{2}{c|}{Supply (V)}             		
				&1.9/1.1/0.9
				&1
				&0.95
				&1.0
				&0.85/1.05	
				&0.8												
				&0.9/0.75					
				&\textbf{1.0}
			\\\hline\multicolumn{2}{c|}{Process (nm)}              
				&28
				&28
				&28
				&65
				&65
				&14			
														
				&4						
				&\textbf{28}
			\\\hline\multicolumn{2}{c|}{Resolution (bit)}					 
				&10
				&8
				&8
				&8	
				&8
				&8			
												
				&10						
				&\textbf{10}
			\\\hline\multicolumn{2}{c|}{Sampling Rate (GS/s)}	 
				&8
				&10
				&20
				&10
				&16
				&10

				&16										
			&\textbf{13}
			\\\hline\multicolumn{2}{c|}{Channels}                          
				&16
				&8
				&16
				&4	
				&8
				&2
														
				&4			
				&\textbf{8$\times$ I/Q}
			\\\hline\multicolumn{2}{c|}{SNDR @Nyquist (dB)}
				&49.0
				&36.9
				&38.9
				&40.1
				&38.1
				&37.2

				&44.5			
				&\textbf{45.8}
			\\\hline\multicolumn{2}{c|}{SFDR @Nyquist (dB)}
				&60.3
				&59.0
				&51.9
				&52.8
				&52.5
				&50.7

				&55.9							
				&\textbf{56.1}
			\\\hline\multicolumn{2}{c|}{\tabincell{c}{Power (mW)}}
				&300
				&21
				&42.1
				&50.8
				&76.2
				&14.8

				&94.2					
				&\textbf{26.4}
			\\\hline\multicolumn{2}{c|}{Area (mm$^2$)}
				&0.184
				&0.022
				&0.042
				&0.095
				&0.22	
				&0.00285$^a$

				&0.008$^a$							
				&\textbf{0.38}
			\\\hline\multicolumn{2}{c|}{FoM$_\mathrm{W}$ (fJ/conv.-step)}
				&162.9
				&37.0
				&29.1
				&61.5
				&72.6
				&24.8

				&42.9					 				
				&\textbf{12.7}
                \\\hline\multicolumn{2}{c|}{FoM$_\mathrm{S}$ (dB)}
				&150.2
				&150.7
				&152.7
				&150.0
				&150.1
				&152.5										
				&153.8					 				
				&\textbf{159.7}
			\\\hline\multicolumn{2}{c|}{\tabincell{c}{Timing-Error\\[-2mm]Calibration-Free?}}
				&no
				&no
				&no
				&no
				&no				
				&no												
				&no											
				&\textcolor{blue}{\textbf{YES}}\
			\\\hline\hline
		\end{tabular}
		\begin{tablenotes}
			\scriptsize
			\item[]$^a$Excluding input/clock distribution network
		\end{tablenotes}
	\end{threeparttable}
\end{table*}


The proposed AFT-based frequency-interleaved ADC has been fabricated in a 28 nm CMOS technology, with the die micrograph shown in \figurename\ref{fig-5.1}.
The ADC occupies an active area of 0.379 mm$^{2}$, among which 0.253 mm$^{2}$ are occupied by the clock and input distribution network.
For testing purposes, the output is decimated by a factor of 4032.
The time-domain signal is reconstructed from the IDFT algorithm.

\figurename\ref{fig-5.2} shows the measured differential nonlinearity (DNL) and integral nonlinearity (INL), which are +1.2/-1 LSB and +1.7/-1.5 LSB, respectively.
\figurename\ref{fig-5.3} shows the measured output spectra with a low-frequency input at 11.34 MHz and a Nyquist input at 6.499 GHz.
At low frequency, the ADC achieves 47.1-dB SNDR and 56.9-dB SFDR.
While at Nyquist, 45.8 dB and 56.1 dB are measured for SNDR and SFDR, respectively. 
In both cases, the interleaving spurs remain buried under the noise floor, demonstrating the inherent robustness to timing errors.

\begin{figure}[t]
	\centering
	\includegraphics[width = 0.93\linewidth]{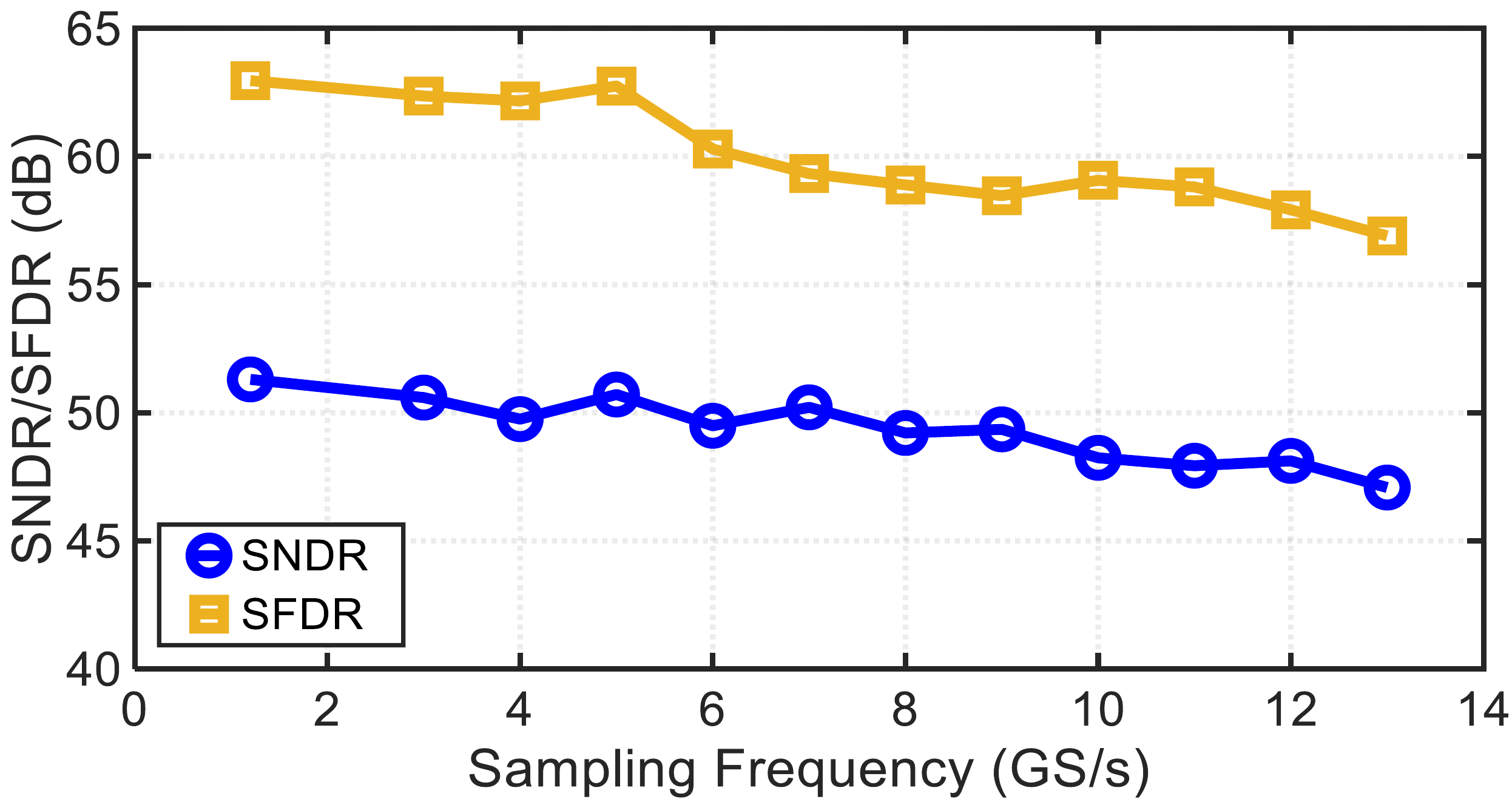}
	\caption{Measured SNDR/SFDR \textit{vs.} sampling frequency.}
	\label{fig-5.4}
\end{figure}

\begin{figure}
	\centering
	\includegraphics[width = 0.93\linewidth]{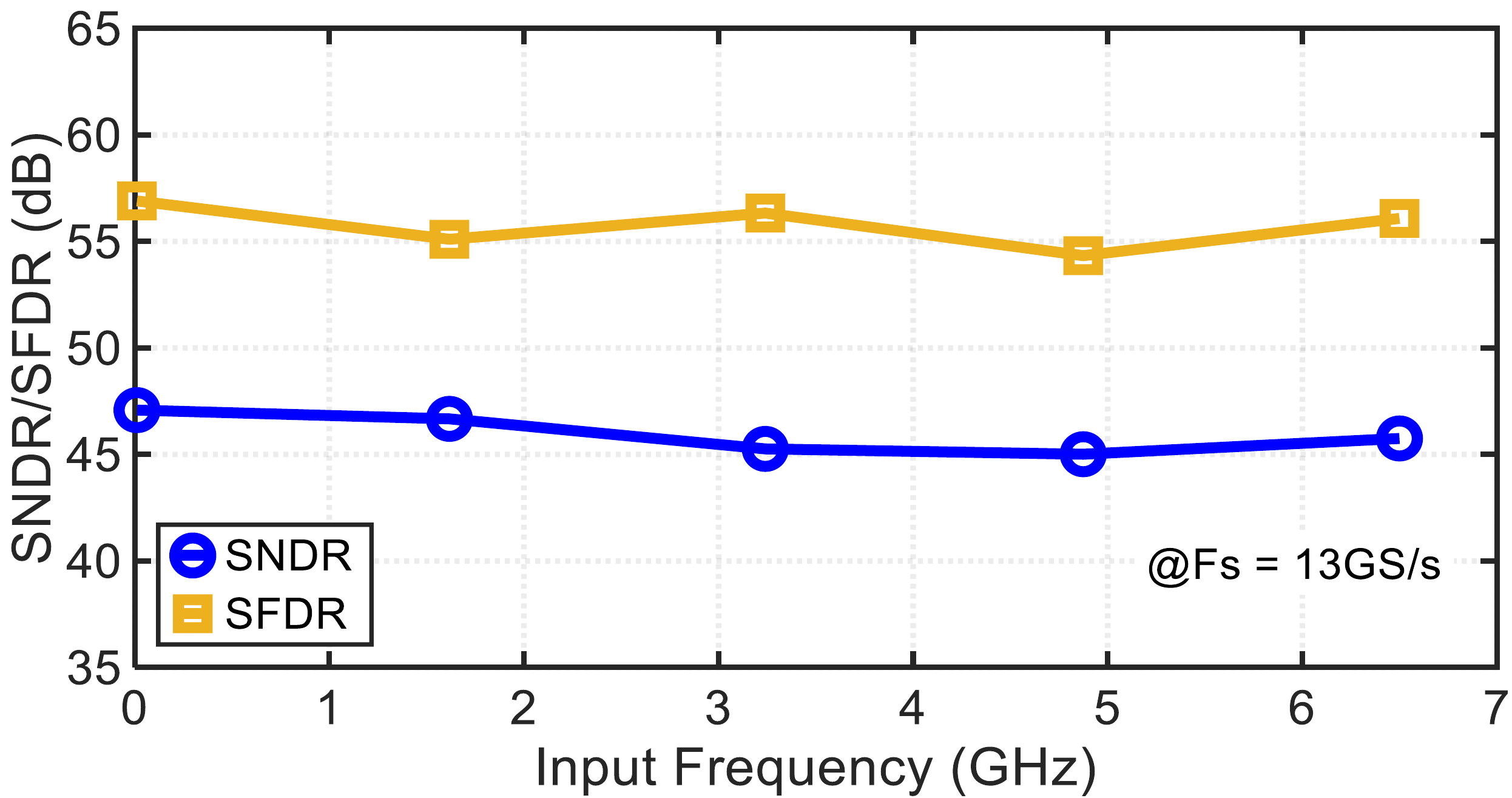}
	\caption{Measured SNDR/SFDR \textit{vs.} input frequency.}
	\label{fig-5.5}
\end{figure}

\figurename\ref{fig-5.4} shows the measured SNDR and SFDR versus sampling frequency, where the SNDR degrades 4.22 dB at 13GS/s, and the measured ERBW is 12 GHz.
\figurename\ref{fig-5.5} shows the measured SNDR and SFDR versus input frequency.
At 13 GS/s, less than 2 dB variation is observed across the measured frequency range.

\begin{figure}
	\setlength{\abovecaptionskip}{0pt}
	\setlength{\belowcaptionskip}{0pt}
	\centering
	\includegraphics[width = 0.99\linewidth]{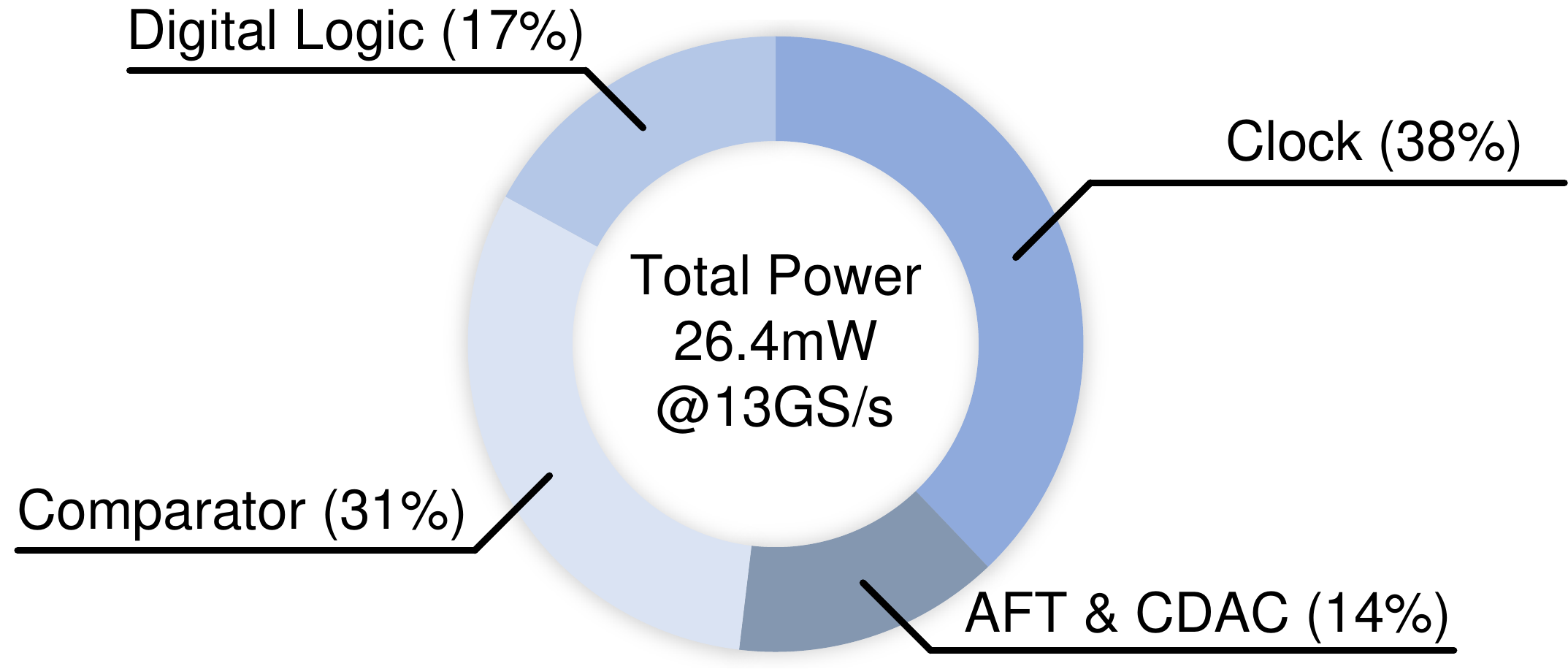}
	\caption{ADC power breakdown.}
	\label{fig-5.6}
\end{figure}

The ADC consumes 26.4 mW under a single 1V supply voltage, resulting in a Walden FoM of 12.7 fJ/conv.-step and a Schreier FoM of 159.7 dB.
\figurename\ref{fig-5.6} shows the power breakdown.
Since each channel of the AFT-embedded sub-ADC has to sample at all phases, a large portion of the total power is consumed by clocks.

\tablename~\ref{table_comparison} summarizes the measured performance of the proposed FI ADC. 
The comparison with state-of-the-art high-speed ADCs is also presented.
With the proposed AFT-based FI architecture, this work demonstrates the first FI ADC prototype that achieves competitive performance compared to the de facto TI implementations.
Furthermore, owing to the inherent timing-error robustness of the FI architecture, the proposed ADC eliminates the need for timing-error calibration, thereby significantly reducing the design complexity and calibration overhead of high-speed ADCs.

%% file: sections/6_conclusion.tex
\section{Conclusion}\label{sec:conclusion}

Timing errors, including timing skew and tracking bandwidth mismatch, are fundamental and require a significant amount of design and calibration effort in high-speed ADCs, where the time interleaving has become the de facto architecture.
The frequency-interleaved architecture offers inherent robustness to timing errors, which can be achieved by the proposed AFT-based approach. 
An 8-point charge-based AFT with lossless implementation of complex-valued irrational coefficients has been proposed, which can be embedded into the CDAC array of the sub-ADC. 
The 28-nm prototype ADC has been measured at 10b 13GS/s, achieving 45.8-dB SNDR and 56.1-dB SFDR at Nyquist, and 12.7 fJ/conv.-step Walden FoM. 
This paper demonstrates the first FI ADC that achieves competitive performance compared to state-of-the-art TI ADCs.
The inherent timing-error robustness of the proposed AFT-based FI architecture, as validated by the calibration-free prototype, establishes a compelling foundation for future high-speed ADC designs.